\documentstyle[twocolumn,pre,aps,psfig]{revtex}

\title{Equilibrium and dynamical properties\\
of two dimensional self-gravitating systems}

\author{Alessandro Torcini$^{1,2}$
\thanks{e-mail: torcini@fi.infn.it}
\thanks{web : http://torcini.de.unifi.it/$\sim$torcini}
and Micka{\" e}l Antoni$^{3}$ 
\thanks{e-mail: antoni@sebigbos.mpipks-dresden.mpg.de}
\\
$^1$ {\small \it Dipartimento di Energetica ``S. Stecco'',
Universit\'a di Firenze, via S. Marta 3, I-50139 Firenze, Italy}\\
$^2$ {\small \it Istituto Nazionale di Fisica della Materia, 
Unit\`a di Firenze, Largo E. Fermi 2, I-50125 Firenze, Italy}\\
$^3$ {\small \it Max-Planck-Institut f\"ur Physik komplexer Systeme,
N{\" o}thnitzer Str. 38, D-01187 Dresden, Germany} \\
}
\author{\parbox{430pt}{\vglue 0.3 cm \small
A system of $N$ classical particles in a 2D periodic cell
interacting via a long-range attractive potential 
is studied numerically and theoretically.
For low energy density $U$ a collapsed phase is identified,
while in the high energy limit the particles are
homogeneously distributed. A phase transition from the 
collapsed to the homogeneous state
occurs at critical energy $U_c$. A theoretical analysis within 
the canonical
ensemble identifies such a transition as first order. But
microcanonical simulations reveal a negative specific 
heat regime near $U_c$.
This suggests that the transition belongs
to the universality class previously identified by
Hertel and Thirring (Ann. of Physics, {\bf 63}, 520 (1970))
for gravitational lattice gas models. The dynamical
behaviour of the system is strongly affected by this 
transition : below $U_c$ anomalous diffusion is observed, 
while for $U > U_c$ the motion of the particles is almost ballistic.
In the collapsed phase,
finite $N$-effects act like a "deterministic" noise source 
of variance ${\cal O}(1/N)$, that restores normal diffusion 
on a time scale that diverges with $N$. As a consequence,
the asymptotic  diffusion coefficient 
will also diverge algebraically with $N$ and superdiffusion
will be observable at any time in the limit $N \to \infty$.
A Lyapunov analysis reveals 
that for $U > U_c$ the maximal exponent $\lambda$
decreases proportionally to $N^{-1/3}$ and
vanishes in the mean-field limit. For sufficiently 
small energy, in spite of a clear non ergodicity 
of the system, a common scaling law 
$\lambda \propto U^{1/2}$ is observed
for any initial conditions. In the intermediate energy range, where 
anomalous diffusion is observed, a strong 
intermittency is found. This intermittent behaviour 
is related to two different dynamical mechanisms of chaotization.
}}
\author{\parbox{430pt}{\vglue 0.3 cm \small PACS numbers:
05.40.+j, 05.45.+b, 05.70.Fh, 64.60.Cn}
}

\date{\today}
\begin{document}
\maketitle

\section{Introduction}

Theoretical \cite{her1} and computational studies \cite{com1,pos1}
have been devoted in the last decades to the study of
thermodynamical properties of $N$-body gravitational Hamiltonian models.
One of the main results is the identification of a
phase transition from a collapsing phase (CP)
observed at low energy to a homogeneous phase (HP) at high energy.
In the CP a fraction of the particles form a
single cluster floating in a homogeneous background of almost
free particles.  Above a critical energy $U_c$
the cluster disappears and all the particles move almost freely.
In the transition region,
the specific heat becomes negative in the microcanonical ensemble.
This instability has been widely studied in astrophysics, where
it is known as the {\it gravothermal catastrophe}
\cite{lyn1,saslaw,lyn2}. A negative specific heat regime 
seems to be thermodynamically inconsistent. But this 
paradox is solved once the non-equivalence of canonical
and microcanonical ensembles in the neighbourhood of such phase 
transition is demonstrated \cite{her1,her2}.
These results have been sucessfully confirmed by numerical
investigations of self-gravitating non-singular 2D systems 
with short range interaction \cite{com1,pos1} and by the study
of a system of particles confined on the surface of a sphere
with variable radius \cite{lyn2}. The reported analysis
were  usually focused on 2D systems, but the features of this
thermodynamical transition are expected to be independent
of the space dimensionality \cite{her1}.

More recently it has been shown that cluster formation 
in simple 1D $N$-body models with long-range attractive interaction
\cite{kon0,ruf} is similar to the {\it Jeans instability}
\cite{jeans} occurring in gravitational systems \cite{ina}. 
In particular, following this analogy Inagaki and Konishi \cite{ina}
were able to derive the correct critical temperature for the 
declustering transition \cite{ruf1}.

Preliminary indications suggesting that clustering phenomena 
can have effects on single particle dynamics have been reported 
for \begin{itemize}

\item (i) a system of fully and nearest-neighbour
coupled symplectic maps with attractive interaction 
\cite{kan1};

\item (ii) atomic clusters \cite{lebastie};

\item (iii) turbulent vortices \cite{cardoso,sol1}.

\end{itemize}
All these systems exhibit a clustered phase associated
to an anomalous diffusion law. The anomalous transport 
is related to the single particle dynamics that intermittently
shows a sequence of localized and almost ballistic behaviours,
corresponding to the trapping and untrapping in 
the cluster, respectively.

Anomalous diffusion has been revealed in dissipative and Hamiltonian models
\cite{geis3,zum2} as well as in experiments \cite{cardoso,sol1}.
The first investigated example of anomalous behaving systems was a 
one dimensional chaotic map proposed
by T. Geisel and co-workers \cite{geis1} that was introduced to describe
enhanced diffusion in Josephson junctions \cite{dub1}. More recently,
anomalous diffusion was also studied in the context of fluid dynamics
and solid state physics both experimentally \cite{sho1,cardoso,sol1} 
and theoretically \cite{geis3,zum2,gri1}. However, the conservative 
models usually 
studied concern the motion of a single particle that evolves in a fixed
two-dimensional periodic potential \cite{geis2,geis3,kla1,fle1,cha1} 
or the dynamics of low dimensional symplectic maps. Only few studies 
have been devoted to extended models with $N >> 1$ \cite{kan1,kan2,kon2}.

The main purpose addressed in this paper is to establish 
a link between equilibrium results, transport properties
and finite $N$-effects for $N$-body Hamiltonian models with
long-range attractive interaction. In order to examine
in detail these points we consider a simplified 2D
self-gravitating model, where the dynamics of each
particle can be written 
in terms of mean-field quantities that are self-consistently defined
through the coordinates of all the particles. Therefore each particle
is moving in a non autonomous self-consistent potential depending
on mean-field quantities.
This peculiarity allows to implement a very efficient code and
therefore to examine systems with a high number of particles
($N \simeq 4,000 - 10,000$) a for long time periods ($t \sim 1 \times
10^6$ - $1 \times 10^7$, where the typical time of evolution
is ${\cal O}(1)$).

We observe a declustering transition, 
analogous to the one studied by Hertel and Thirring, and, in 
correspondence of the CP, anomalous diffusion is 
detected. Above the transition point also the dynamical behaviour 
changes and the particles move almost freely.

In the CP, the shape of our self-consistent single particle 
potential resembles that of the so called egg-crate
potential. Several studies were devoted to the
motion of a single particle in an autonomous egg-crate
potential, i.e. when the potential represents 
a fixed landscape for the particle 
\cite{geis3,zum2,geis2,kla1,fle1,cha1}.
In particular, anomalous diffusion was observed
and such behaviour was explained as
due to trapping and untrapping of the particle orbit 
in a self-similar hierarchy of cylindrical cantori \cite{geis3}. 
As a result the trajectory of the particle on the 2D
surface is similar to a L{\'e}vy walk. This suggests
that anomalous transport, present in our model, should
be explained by similar mechanisms.

Due to the self-consistent nature of the potential in the model
we consider, finite $N$-effects induce a time fluctuation in the potential
seen by a single particle and essentially act as a noise 
source of typical intensity $O(1/\sqrt{N})$. Environmental 
fluctuations are known to restore normal diffusion on long time 
scales  when added to a dynamical systems exhibiting anomalous 
diffusion \cite{flo1,bet1}. Indeed, this is what happens also in
our model, apart that now the fluctuations are intrinsically related
to its deterministic dynamics and not to an external
bath. In the mean-field limit the fluctuations disappear
and anomalous diffusion is present at any time.

The study of the degree of chaoticity in the present model
reveals that the system (for low energy) is highly non-ergodic
and, despite the fact that thermodynamical properties seem not to depend on
initial conditions, the dynamical indicators (e.g. the maximal
Lyapunov exponent) are heavily affected from the initial state of 
the system. Despite this non-ergodicity,
the low energy dependence of the maximal Lyapunov
shows a common scaling law for all the considered initial conditions.
In the intermediate energy range, two distinct chaotic mechanisms 
are identified in our model and give rise to intermittency in the 
dynamical evolution. For high energy the system 
becomes integrable in the mean-field limit and the Lyapunov
vanishes as an inverse power law of $N$.

This article is organized in the following way. In section II, we 
present the Hamiltonian model we focus on. 
In section III, we report a thermodynamical description of the model
within the canonical ensemble. Equilibrium values of the mean-field are
derived and the existence of a phase transition from
a homogeneous to a collapsed phase is shown. Section IV is
devoted to numerical description of dynamical properties. In 
particular, we'll concentrate on the diffusion of single 
particles and on the influence of
finite $N$-effects. The chaotic behaviour of the model is 
discussed in section V in a broad energy range.
Finally, in section VI the reported results are briefly discussed.

\section{The Model}

In the present paper we study the static and 
dynamical properties of a $N$-body system enclosed
in a $2$-D periodic cell, with a classical long-range 
inter-particle potential. 
The dynamics of each particle is ruled by the Hamiltonian :
\begin{eqnarray}
H&& = \sum_{i=1}^N \frac{p_{i,x}^2+p_{i,y}^2}{2}
+\frac {1}{2N} \sum_{i,j}^N
\Biggl[3- \cos(x_i-x_j)
\nonumber \\
&&-\cos(y_i-y_j)-\cos(x_i-x_j)\cos(y_i-y_j)\Biggr] = K+V \quad ,
\label{ham}
\end{eqnarray}
where $(x_i,y_i) \in [-\pi,\pi] \times [-\pi,\pi]$,
$(x_i,p_{i,x})$ and $(y_i,p_{i,y})$ are the two couples of conjugate
variables. The particles are assumed to be identical and to have 
unitary mass. $K$ (resp. $V$) is the kinetic (resp. potential) energy. 
The  reference energy is chosen in such a way that the energy 
of the system vanishes when all the particles have the same position 
($V=0$) and zero velocity ($K=0$). The presence of the third term in 
the potential energy is essential in order to ensure an interaction 
between the two couples of conjugate variables.

The considered interparticle potential belongs to the following class 
of $2-$D periodic potentials :
\begin{eqnarray}
V&& =\sum_{i=1}^N V_s({\bf r}_i) \quad {\rm with} 
\nonumber \\
&& V_s({\bf r})=\frac 1 {4N} \sum_{j=1}^N \sum_{0 < {\bf k}^2 \le s} 
c(k) \Bigl[1-\cos[ {\bf k} \bullet ({\bf r}-{\bf r}_j)]\Bigr] 
\label{pot1}
\end{eqnarray}
where $s$ is a parameter that determines the number of harmonics 
included in the Fourier expansion of $V$, ${\bf r}=(x,y)$ and 
${\bf k}=(n_x,n_y)$ is the wave-vector, with $n_x$ and $n_y$ 
two integers.  
$c(k)=c(-k)$ is a real valued function of 
the modulus of ${\bf k}$, that fixes the coupling 
strength of the harmonic $k$.
For model (\ref{ham}), $c(k)=1/k^2$ and $s=2$. For the 
same choice of $c(k)$, but in the limit $s \to \infty$, 
a $2-D$ self-gravitating Newtonian potential 
$V_\infty ({\bf r}) \propto \sum_{i=1}^N{\rm log}
|{\bf r}-{\bf r}_j|$ is recovered, once a rescaled
time $t/\sqrt{N}$ is considered.
It should also be noticed that interactions of the type 
$\log|r|$ among particles arises also in point vortices model 
for 2D turbulence \cite{fri1}. Therefore the simplified model
(\ref{ham}) should share some common behaviour 
with point vortices systems.

The interparticle potential appearing in (\ref{ham}) 
is essentially the Fourier expansion of $V_\infty ({\bf r})$
limited to its first three terms and the corresponding
equations of motion for the coordinates $x_i$ are 
\begin{equation}
\ddot x_i= -\frac 1 N
\sum_{j=1}^N\Biggl[\sin(x_i-x_j)+\sin(x_i-x_j)\cos(y_i-y_j)\Biggr] 
\label{eqmv}
\end{equation}
due to the symmetry of (\ref{ham}), the equations of motion for 
$y$ are obtained exchanging $x \leftrightarrow y$.

Previous investigation of model (\ref{ham}) \cite{ant1}
has revealed that at low energy $U=H/N$ the particles,
due to attractive coupling, are organized
in a unique collective structure ({\it cluster}) with 
$U$-dependent spatial extension. This clustered phase 
(CP) survives up to a critical energy $U_c \approx 2$.
Above $U_c$ a phase transition occurs to a homogeneous phase (HP). 
This transition will be discussed in more detail in the next 
section. Here, we will limit to evidence collective behaviours in the
system by rewriting model (\ref{ham}) in terms of 
the following mean-field variables :
\begin{eqnarray}
{\bf M}_1&& = (<\cos(x)>_N,<\sin(x)>_N) 
\nonumber \\
&& = M_1 \enskip ( \cos(\Phi_1),\sin(\Phi_1) )
\label{meanfield1}
\end{eqnarray}
\begin{eqnarray}
{\bf M}_2&& = (<\cos(y)>_N,<\sin(y)>_N) 
\nonumber \\
&&= M_2 \enskip ( \cos(\Phi_2),\sin(\Phi_2) )
\label{meanfield2}
\end{eqnarray}
\begin{eqnarray}
{\bf P}_1&& = (<\cos(x + y)>_N, <\sin(x +  y)>_N) 
\nonumber \\
&&= P_1 
\enskip 
( \cos(\Psi_1),\sin(\Psi_1) )
\label{meanfield3}
\end{eqnarray}
\begin{eqnarray}
{\bf P}_2&& = (<\cos(x - y)>_N, <\sin(x - y)>_N) 
\nonumber \\
&&= P_2 \enskip
( \cos(\Psi_2),\sin(\Psi_2) )
\label{meanfield4}
\end{eqnarray}
where $<..>_N$ denotes the average over all the particles in
the system; $\phi_z$ (resp. $\psi_z$) and $M_z$ (resp.
$P_z$) are the argument and the modulus of the mean-field 
vectors ${\bf M}_z$ (resp. ${\bf P}_z$) with $z=1,2$. 
The moduli $M_z$ and $P_z$ are maximal and equal to 1
when all the particles have the same position and
their value decreases when the spatial distribution of the 
particles extends. For an homogeneous 
distribution, due to finite $N$ effects and according
to the central limit theorem, $M_z \approx P_z \approx 
O(1/\sqrt{N})$ for $z=1,2$. Therefore,
the quantities $M_z$ and $P_z$ can be thought as 
order parameters characterizing the degree of clustering 
of the system \cite{kan1,ruf}.

By reexpressing the equation of motion (\ref{eqmv}) for both 
coordinates $x$ and $y$ in terms of the mean-field variables $M_z$
and $P_z$, one straighforwardly shows 
that the time evolution of each particle $i$ is ruled by 
the following single particle non-autonomous Hamiltonian :
\begin{eqnarray}
h_i&& = \frac{p_{x,i}^2+p_{y,i}^2} 2 + \Bigl[3-M_1\cos(x_i-\phi_1)-
M_2\cos(y_i-\phi_2)
\nonumber \\
&&-\frac 1 2 P_1\cos(x_i+y_i-\psi_1)-
\frac 1 2 P_2\cos(x_i-y_i-\psi_2) \Bigr] =
\nonumber \\
&&= K_i + V_i
\label{ham1}
\end{eqnarray}
\noindent
where $K_i$ and $V_i$ are the 1-particle kinetic and
potential energy, respectively. Hence, each particle moves 
in a mean field potential $V_i$ determined by the instantaneous 
positions of all the other particles of the system.

Since $V$ is invariant under the transformations 
$x \leftrightarrow -x$, $y \leftrightarrow -y$ and 
$x \leftrightarrow y$, it turns out that in the 
mean-field limit ($N\to \infty$ with constant
$U=H/N$), $M_1=M_2=M$ and $P_1=P_2=P$. Moreover, 
in this limit and assuming 
$\phi_z=\psi_z=0$, the $1-$particle potential is similar to 
the egg-crate potential studied in references
\cite{geis2,geis3,kla1,fle1,cha1}.
In Fig. (\ref{fig1}) the shape of $V_i$
at a given time for $U = 1.00$ and 
$N=4,000$ is shown.
The potential is periodic along the two spatial
directions and in each elementary cell it has 4 maxima 
($V_M=3+2M-P$), 4 saddle points ($V_s=3+P$) and a
minimum ($V_m=3-2M-P$). The depth of the potential well
is $\simeq (V_s-V_m)$
and the center of the cluster will coincide
with the position of the potential's minimum.
In the limit $U \to 0$,
$M$ and $P \to 1$ and therefore $V_M \to 4$,
$V_s \to 3$ and $V_m \to 0$. In this limiting
situation all the particles are trapped 
in the potential well they create.
For increasing $U$, the kinetic contribution
becomes more relevant and the number of clustered
particles reduces. This implies that also
the values of $M$, $P$ and of the well depth 
decrease. For $U \ge U_c$, the particles are
no more clustered and in the
limit $N \to \infty$ the single particle potential
becomes flat $V_M = V_s = V_m = 3$ and time independent.

However, due to finite $N$ effects, the instantaneous mean-field 
variables fluctuate with typical time $1$ within the statistical 
band range $\sim 1/\sqrt{N}$. 
This implies that in the $CP$ the single particle will move 
in a fluctuating potential landscape. 
As a consequence, the system admits 
a time pulsating separatrix that
sweeps a phase space 
domain proportional to $1/\sqrt{N}$ in the neighborhood of the
average position of the separatrix.

Due to the shape of $V_i$, we can
qualitatively distinguish at each time three sub-groups 
of particles:
the low energy trapped particles (LEP) oscillating in the 
self-consistent well of the potential (with energy $h_i \sim V_m$),
the intermediate energy particles (IEP) evolving in the stochastic sea 
neighbouring the separatrix ($h_i \simeq V_S$) and the high energy 
particles (HEP) moving almost freely ($h_i >> V_M$). Obviously,
during the time evolution
each particle can pass from one sub-group to another.

\section{Thermodynamical Aspects}

\subsection{Canonical Prediction for the Equilibrium Properties}

The problem of finding the thermodynamical potentials of model (\ref{ham})
in the canonical ensemble is resolved starting as usual from the partition 
function $Z(N,\beta)$, where $\beta$ is the inverse temperature. 
For Hamiltonian (\ref{ham}), the partition function factorizes in 
$Z(N,\beta) = Z_K \times  Z_V$, where~:
\begin{equation}
Z_K = \int_{{I\!\!R}^{2N}} \exp
	    \left(- \beta \sum_{i=1}^N \frac {p_{x,i}^2+p_{y,i}^2} 
2\right) {d^N}p_x {d^N}p_y 
  = \left(\frac {2 \pi} \beta \right)^N
\label{eqZKin}
\end{equation}
is the kinetic part obtained assuming a Maxwell-Boltzmann
distribution for the momenta.
The potential contribution $Z_V$ is evaluated by considering the
expression of the potential $V$ in terms of mean-field variables 
${\bf M}_z$ and ${\bf P}_z$. 
As already mentioned, in the mean-field limit due to the symmetries 
of $V$ we have $M_1=M_2=M$ and $P_1=P_2=P$. Consequently, $V$ can 
be reexpressed as a quadratic form in $M$ and $P$ and the resulting 
expression for $Z_V$ reads \cite{ruf}:
\begin{equation}
Z_V \propto \int_{S^{2N}}
      \exp \left[\beta \frac N 2 \left (2 M^2+P^2\right) \right]
\quad d^{N}x \quad d^{ N}y
\label{eqZpot}
\end{equation}
\noindent
where $S$ stands for the unit circle and where the constant part of 
$V$ is omitted. We can evaluate the integral in
Eq. (\ref{eqZpot}) by using the Hubbard-Stratonovich trick:
\begin{equation}
\exp (c A^2) \propto \int_{{I\!\!R}^2}
      \exp (-u_A^2 + 2 \sqrt{c} {\bf{u}}_A\cdot{\bf{A}}) {d^2}\bf{u}_A
\label{eqGauss}
\end{equation}
where ${\bf A}$ is a two component real vector and $c$ a 
positive constant. Following \cite{ruf,els2}, we substitute
expression (\ref{eqGauss}) into Eq. (\ref{eqZpot}) with
($A=M,c=\beta N$) and ($A=P,c=\beta N/2$) and perform the rescaling 
$\sqrt{2\beta/N} {\bf u}_w \rightarrow {\bf u}_w$ for $w=M,P$.
Then expressing ${\bf u}_w = (u_w \cos \theta_w, u_w \sin \theta_w)$ 
in polar coordinates, we obtain:
\begin{eqnarray}
Z =&& Z_K \times  Z_V \propto 
\nonumber \\
&&\int_{{(I\!\!R \times S)}^2} 
u_M u_P \exp (-N G) \quad du_M d\theta_M du_P d\theta_P 
\label{eqZpot2}
\end{eqnarray}
\noindent
where $G = (u_M^2+u_P^2)/{\beta} - {\rm log} R + {\rm log}
\beta $ is a strictly positive real valued function independent of 
$N$ and $R$. The main contribution to $Z$ in the limit 
$N~\rightarrow~\infty$ can be evaluated with the saddle point
method, and reduces to the estimation of the minimum
of the function $G$. The minimum value will not depend on
$\theta_M$ and $\theta_P$, because $R$ is maximal
with respect to these two variables  for 
$(\theta_M,\theta_P)=(0,0)$ for any $u_M$ and $u_P$.
Its expression thus reduces to:
\begin{eqnarray}
&&R_m(u_M,u_P) =\max_{\theta_M,\theta_P} R(u_M,u_P,\theta_M,\theta_P) =
\nonumber \\
&& =2\pi\int_0^{2\pi} I_0\Bigl(u_M+\sqrt{2}u_P\cos(s)\Bigr)
\exp\Bigl(u_M \cos(s)\Bigr)ds
\label{defH}
\end{eqnarray}
\noindent
where $I_0$ is the zero order modified Bessel function. 
Finally, the minimum of $G = G(\beta, u_M, u_P)$ is obtained
numerically using expression (\ref{defH}) and a Raphson-Newton 
scheme. The $\beta$ dependence of the solutions 
$({\bar u}_M,{\bar u}_P)$ for which $G$ is minimal is reported 
in Fig. \ref{fig2}. At large $\beta$ (low temperature) the 
minimum is 
located far away from the origin and corresponds 
to the CP i.e. to the broken symmetry phase.
When $\beta=2$, a saddle-node bifurcation occurs at the origin 
$({\bar u}_M,{\bar u}_P)=(0,0)$ (corresponding to the HP), 
where a second minimum appears. 
This can be shown with a Taylor expansion of $G$ in the neighborhood 
of the origin :
\begin{eqnarray}
G(\beta,u_M,u_p)&& 
\approx u_M^2\Biggl(\frac 1 \beta - \frac 1 2\Biggr)
+u_P^2\Biggl(\frac 1 \beta 
- \frac 1 4\Biggr) 
\nonumber \\
&&-{\rm log}\Bigl(4\pi^2\Bigr)
+ \ln \beta
\label{tayG}
\end{eqnarray}
From the above equation, one can observe that for $\beta > 2$ the 
origin is a saddle point (it is a local maximum if $\beta > 4$) while 
for $\beta < 2$ the origin is a local minimum. The upper inset of 
Fig. {\ref{fig2} shows the details of the temperature 
domain where both minima coexist. 
When decreasing $\beta$,
a second bifurcation occurs for $\beta = 1.81$ and the 
minimum corresponding to the 
broken symmetry disappears while the one at the origin persists. 

The mean-field variables (\ref{meanfield1}),(\ref{meanfield2}),(\ref{meanfield3})
and (\ref{meanfield4}) can be estimated from their joint characteristic function:
\begin{equation}
\Psi({\bf \sigma_M},{\bf \sigma_P})
=<e^{({\bf \sigma_M \bullet M + \sigma_P \bullet P})}>
\label{jpf}
\end{equation}
where the average $< \cdot >$ runs over the corresponding Gibbs
ensemble. The quantities ${\bf \sigma_M}$ and ${\bf \sigma_P}$ can 
be considered as external fields while ${\bf M}$ and ${\bf P}$ are
the equivalent of magnetizations or polarization vectors.
The function (\ref{jpf}) can be explicitly computed
and its first momenta are the moduli of the mean-field variables :
\begin{eqnarray}
M &&= \frac {\partial \Psi}{\partial |\sigma_M|}  =
\frac {2 \pi}{R_m({\bar u}_M,{\bar u}_P)}  \times
\nonumber \\
&& \int_{S}
ds \cos(s) I_0 \Bigl({\bar u}_M+\sqrt{2}{\bar u}_P\cos(s)\Bigr)
\exp\Bigl[{\bar u}_M \cos(s)\Bigr]
\label{defMx2}
\end{eqnarray}

\begin{eqnarray}
P &&= \frac {\partial \Psi}{\partial |\sigma_P|} =
\frac {2 \pi} {R_m({\bar u}_M,{\bar u}_P)}  \times
\nonumber \\
&&\int_{S}  ds \cos(s)
I_1\Bigl({\bar u}_M+\sqrt{2}{\bar u}_P\cos(s)\Bigr)
\exp \Bigl[{\bar u}_M\cos(s)\Bigr]
\label{defMxy2}
\end{eqnarray}
where function $R_m$ was introduced in Eq. (\ref{defH}), $I_1$ is the first order
modified Bessel function. From these expressions it is easy to check that
when $({\bar u}_M,{\bar u}_P)$ coincide with the origin, then $M=P=0$.
This allows us
to identify such solution with the HP. While if $({\bar u}_M,{\bar u}_P)$
is located away from $(0,0)$ the solution corresponds to the CP.

More specifically, $M$ and $P$ can be considered as two order parameters 
that describe the spatial extention of the cluster in the $(x,y)$ plane 
(i.e. the degree of clustering of the system). 
The analytical estimation for $M$ and $P$ are plotted in
Fig. \ref{fig3}(a) (solid line) as functions of $U$. 
For $U \to 0$, $M$ and $P$ tend to $1$ as expected for a fully 
ordered phase. 
For increasing $U$, kinetic effects increase, the size of the cluster 
thus grows and consequently $M$ and $P$ decrease.
For $1.6 < U < 2.0$ the analytical values are not reported, because
in this energy range it is impossible to estimate them with a canonical 
approach. This simply reflects the first order nature of the 
transition. For $U > 2.0$, $M=P=0$ because the system is no more 
exhibiting a clustered phase.

However, within a canonical ensemble it is more appropriate to 
analyze  $M$ and $P$ as a function of the temperature $T = \beta^{-1}$.
Both these quantities decrease with $T$ and vanish when 
$ T >  T_c \sim 0.55$,
their dependence from $(T_c - T)^{-1}$ is shown in Fig. \ref{fig3}(b).
Close to the critical temperature both $M$ and $P$ exhibit
a jump indicating a first order transition between $CP$ and $HP$.

Thermodynamic potentials can be straighforwardly obtained 
from the partition function~:
\begin{eqnarray}
F&& = - \lim_{N \rightarrow \infty} \Bigl({\frac 1 {N \beta}} \log Z\Bigr)
=\frac {G(\beta,{\bar u}_M,{\bar u}_P)} \beta, \quad
\nonumber \\
 U&& = \frac {\partial (\beta F)} {\partial \beta}=
\frac {\partial G(\beta,{\bar u}_M,{\bar u}_P) } {\partial \beta}
\label{defFU}
\end{eqnarray}
where $F=F(\beta)$ is the Helmholtz free energy and $U$ the specific 
(or internal) energy. In order to find the stable configuration
of the system we should look for the absolute minimum of $F$. 
The inverse temperature dependence of the minimum of $F$ is reported 
in Fig. \ref{fig4}(a). In the limit $\beta \to 0$ the free
energy will have a minimum at the origin 
$({\bar u}_M,{\bar u}_P) = (0,0)$ and the HP will be observed,
while in the limit $\beta \to \infty$ the minimum will
be located at $ 0 < {\bar u}_P < {\bar u}_M$ and the system
will be in the CP.
In the range $\beta \in [1.81,2.0]$,
$F$ displays two coexisting minima, that we denote, with  
straightforward notation, as $F_{HP}$ and $F_{CP}$. However for
$\beta > \beta_c = 1.84$ the CP is observed, because
$F_{CP} < F_{HP}$, while for $\beta < \beta_c$ the HP
prevails since $F_{CP} > F_{HP}$. At $\beta = \beta_c$ the 
two minima are equivalent ($F_{CP} = F_{HP}$),
this is a further indication that the transition is of first order.

Fig. \ref{fig4}(b) represents the temperature 
$T$ as a function of the internal energy $U$. For small energies,
$U \approx T/2$ (i.e. there is a virial) and the particles are all trapped 
in a single cluster.  For $U > U_c \approx 2$, the system is in $HP$ and 
$T$ is also increasing linearly with $U$. This indicates that the system 
behaves like a free particle gas. These two regimes correspond to the two 
integrable limits of model (\ref{ham}). 

In the intermediate energy range, the tendency of 
the system to collapse is balanced by the kinetic energy and $T$
is no more proportional to $U$. For $1.6 < U  < 2$, the amount of 
kinetic energy is such that a significant fraction of particles escape 
from the cluster. The system is then characterized by the presence of 
two coexisting phases: the first is supported by the fraction of 
LEP that are trapped in the well of the self-consistent potential and 
the second is due to the HEP that have large kinetic energy and thus 
behave almost freely.

\subsection{Microcanonical Results : Numerical Findings}

In order to obtain microcanonical results,
standard molecular dynamics (MD) simulations have been
performed for the present model (\ref{ham}) 
within a NVE ensemble \cite{allen}.  
The equations of motion are integrated starting from 
their formulation  in terms of the single 
particle Hamiltonian $h_i$ (\ref{ham}), where the 
dependence of the dynamics of each particle from 
mean-field variables is made explicit. 
This approach turns
out to be quite efficient from a computational point of 
view, because the CPU time increases only linearly
with $N$ (and not proportionally to $N^2$ as usual).
This allows us to perform 
simulations of the model (\ref{ham}) for high numbers
of particles (up to  $N = 10^4$) and for quite long
integration time (up to $t= 5 \times 10^7$).
As integration scheme we have adopted a recently developed 
$4^{th}$ order simplectic algorithm \cite{mla1}:
with an integration time step $dt=0.3$ relative energy fluctuations 
$\Delta U / U$ remains
smaller than $10^{-8}$ for any considered $N$ and $U$. 

In the simulations, the particles are
initially clustered in one single point and
with a "water-bag" distribution for the velocities.
Other initial conditions, for example,  with particles
uniformly distributed over the box and with Maxwellian
velocity distribution have been tested. But as far as 
thermodynamical properties were concerned no significative
differences have been observed.
In order to avoid transient effects,
the reported thermodynamical and 
dynamical quantities have been measured starting 
from initial "relaxed" states resulting from 
sufficiently long preliminary simulations.
Moreover, the averages have been typically performed
over total integration times $t \simeq 10^6$.
In the present section, the thermodynamical properties
will be considered, while the dynamical ones will be
the subject of the two following sections.

We define the measured temperature $T$, following
the equipartition theorem, through the time averaged specific 
kinetic energy $T=<K(t)/N>_t$. In Fig. \ref{fig4}(b), the measured 
$U$-dependence of $T$ is reported. The MD-estimation of $<M>_t$ 
and $<P>_t$ are shown  in Fig. \ref{fig2}(a). 
Both figures indicate that the numerical findings are in good 
agreement with the canonical predictions, apart in the interval
$1.6 < U < U_c$, where canonical estimate are not available.
In this energy range (enlarged in the inset of Fig. \ref{fig3}(a))
we observe a decrease of $T$ with $U$. This indicates that
the specific heat will be negative for energies slightly below
$U_c$. The existence of such a negative specific heat regime 
in proximity of the "declustering" transition for gravitational
potential was predicted in 1971 by Hertel and Thirring 
\cite{her1}. These authors studied a simple classical cell
model and noticed the non-equivalence of canonical and
microcanonical ensembles in this region. These predictions 
have been successively confirmed by numerical findings 
for short ranged non-singular attractive potential 
\cite{com1,pos1}. 
Negative specific heat regimes are forbidden within the Gibb's canonical
ensemble, because physically unstable, and thus they are bridged by a 
constant temperature line (that in our case is $T_c \approx 0.54$). 
This picture is similar to the Maxwell construction for the van der Waals
isotherms in the liquid-vapor coexisting region.
This prohibition does not hold in the microcanonical ensemble and the 
physical implication of this is fundamental in astrophysics where 
negative specific heat regimes have been studied for several decades
\cite{lyn1}.

Within the microcanonical context, the phenomenon of negative specific 
heat can be understood following a heuristic argument \cite{her1}. 
As shown in Fig. \ref{fig3}(a) approaching the transition
the values of $M$ and $P$ decay very fast. 
This suggests that a very limited increase of $U$ 
yields a significant reduction of the number of LEP and 
thus a strong increase of the potential energy. 
Total energy being conserved, this excess of potential 
energy has to be compensated by a loss of kinetic energy,
as a result the system will become cooler \cite{her1}.

From our numerical investigations it turns out that for any $U$ 
the average thermodynamical quantities (namely, $T$,$M$ and $P$)
are independent from $N$ (for $100 \le N \le 10,000$),
apart at the critical energy. At $U = U_c$, a clear
decrease of $M$ is observed from a value 0.198 at $N=200$
to 0.063 at $N=10,000$ indicating that the system is
approaching the HP (a similar behaviour is observed 
for $P$). This is confirmed from the fact that also the 
temperature tends to the corresponding HP value
($T= U - 3/2$) for increasing $N$: $T$ varies from 0.55
at $N=200$ to 0.51 at $N=10,000$. This effect is 
probably due to the strong fluctuations in the single
particle potential $V_i$ due to finite $N$-effects,
that become dramatic at the transition. For small
$N$-values a clustered situation seems to be favorite,
while in the mean-field limit the HP is finally recovered.

\section{Dynamical Properties}

\subsection{Dynamical transition and transport properties}

Once examined the microcanonical thermodynamical results, we now 
concentrate on the description of the transport properties of 
(\ref{ham}). In the $CP$ each particle of our system moves
in a single particle potential $V_i=V_i(t)$ ({\ref{ham1}) 
that instantaneously is similar to the so-called egg-crate 
potential (see Fig. \ref{fig1}) \cite{geis2,geis3,kla1,fle1,cha1}. 
The study of single particle motion 
in egg-crate fixed potential was originally motivated in the 
context of adatoms diffusion over a rigid surface \cite{sho1}.
For a single particle moving in such fixed potential the self dynamics 
is known to be anomalously diffusive when the particle is channeling, 
{\it i.e}. when its energy 
lies between $V_M$ and $V_s$ \cite{geis2,geis3,kla1,fle1,cha1}. 
This anomalous behaviour is due to the competition
of laminar and localized phases. Indeed, channeling 
particles intermittently show an almost ballistic motion 
along the channels of the potential interrupted by localized sequences, 
where the particles bounces back and forth on the maxima of the 
potential.

Usually anomalous diffusion  has been studied in 
systems with very few degrees of freedom \cite{zum2,geis3}.
Only few attemps have been made to consider $N$-body 
dynamics \cite{kan1,kan2,kon2}. One of the reasons of this is that
the theoretical analysis of dynamical properties of high 
dimensional Hamiltonian systems is particularly complicated
and not fully understood. Moreover, accurate simulations
of such systems for $N >> 1$ and over long time intervals, 
necessary  to understand the asymptotic diffusive regime, 
are quite difficult to perform.

In this section we first focus on the numerical description
of the transport properties of (\ref{ham}) and we will try to give 
some indications relative to the basic dynamical mechanisms 
governing single particle transport.
To this aim, we consider the time dependence of the mean square 
displacement (MSQD), that usually reads as
\begin{equation}
<r^2(t)> \propto t^\alpha
\label{eqMSQD}
\end{equation}
where the average $<.>$ is performed over different time origins and 
over all the particles of the system. 
The transport is said to be anomalous when $\alpha \ne 1$: 
namely, it is subdiffuse if $0 < \alpha < 1$, superdiffusive 
if $1 < \alpha < 2$ and ballistic for $\alpha = 2$ \cite{geis3,gri1,zum2}. 
The usual Einstein diffusion law corresponds to $\alpha=1$
and in 2D can be written as $<r^2(t)> = 4 D t$, where $D$ is the 
self-diffusion coefficient.
We consider in this paper, $D$ and $\alpha$
as the basic relevant observables for the description of transport.

More refined diagnostics might be considered, for example the 
probability distribution of the time intervals whithin which the 
trajectory of the particle remains trapped.
The latter was explicitely computed both for simple maps 
and for a single particle moving in a fixed egg-crate potential
landscape and shown to exhibit a power-law decay
\cite{geis3,zum2,geis2,zum1,geis4} responsible of the 
anomalous diffusion \cite{geis2,kla1}.
In the limit $N \to \infty$ we expect to find similar 
indications also for model (\ref{ham}) once the IEP's
dynamics is considered.
But for finite $N$, the energy of each particle will not be constant
in time. Therefore a particle initially of type IEP can become
a LEP or a HEP, giving rise to a much more complicated 
dynamical behaviour, that cannot be simply described 
in terms of the localization time distribution. However, if one 
considers an initially IEP and follows its trajectory for some time
it displays features quite similar to the so-called 
Levy walks \cite{zum2} (see Fig. \ref{fig5}). This kind of 
trajectories has been usually identified when anomalous diffusion occurs.
Therefore, we believe that also for model (\ref{ham})
anomalous diffusion will be observable.

As we already mentioned, finite $N$-effects play a determinant role in the 
dynamics. Due to self-consistency, they are responsible for the  
fluctuations in time of the mean-field quantities $M_{1,2}$ and $P_{1,2}$. 
The potential experienced by each particle thus fluctuates
in time and particles having an energy close to $V_s$ have the possibility to 
be trapped in the potential well as well as to escape from it.
This implies that the localization phenomena illustrated in Fig. \ref{fig5} 
are not only due to bounces of the particle on the maxima of the potential,  
but also to trapping in the potential well due to separatrix crossing.

We have argued from simple considerations that diffusion should be
anomalous in $CP$. This is indeed the case, as confirmed from
direct evaluation of the MSQD in a quite extended energy interval.
 An example of this is reported
in  Fig. \ref{fig6} (a) for $U=1.1$ and $N=4,000$.
The diffusion is anomalous for times smaller than a crossover time 
$\tau$ beyond which the Einstein's diffusion law is recovered 
$<r^2(t)> = 4 D t$.  A similar behaviour for the MSQD was previously 
observed for a system of $N$ 
simplectic (globally and locally) coupled maps \cite{kan2,kon2}, but with 
a subdiffusive (i.e. with $\alpha < 1$) short time dynamics.

The direct study of the velocity autocorrelation function (VACF) 
$Z(t)$ confirms the general features seen for the MSQD in the CP: 
namely, on times $t < t_v$ the VACF is characterized by a
long-time tail that decays as $t^{\alpha-2}$.
This power-law decay is fully consistent with the corresponding 
one observed for the MSQD \cite{gri1}. For times longer than $t_v$, 
the VACF decreases exponentially as usually expected for
brownian motion (see Fig. \ref{fig6} (b)).
It is therefore reasonable to expect that $\tau \propto t_v$.

The energy dependence of the $\alpha$-values 
is illustrated in Fig. \ref{fig7}. 
It shows up clearly that the thermodynamical phase transition 
from $CP$ to $HP$ is associated to a dynamical transition from 
superdiffusion (with $1.3 < \alpha < 1.9$ for $0.4 \le U < 2.0$)
to ballistic motion (with $\alpha \simeq 2$ for $U \ge U_c$).
In the $CP$ regime, we observe an increase of $\alpha$ from
$1.3 \pm 0.1$ to $1.9 \pm 0.1$, that is due to the
modification of the shape of the single particle potential.

It is reasonable to expect that this enhanced diffusive behaviour 
with an energy dependent exponent 
is linked to the fraction of channeling particles. The number of such 
particles is related to the depth of the potential well 
(namely, to $V_M - V_m$) and to the energy width of the channels 
$V_M-V_s$. At very low energy $U < 0.3$, all the particles are essentially
trapped in the potential well and none is channeling: no diffusion 
is observed in this case.
For increasing energy a fraction of particles (due to the decrease
of $V_M - V_m$) will escape from the well and some of them
get enough energy to move along the channels: anomalous diffusion
is then evidenced.  The increase
in the value of the exponent $\alpha$ is due to the fact that also
the channel width $V_M-V_s$ grows with $U$ (see Fig. \ref{fig7}).
However, for $U$ approaching $U_c$ the number of untrapped 
particles increases noticeably, but now the channel width vanishes 
abrubtly this implies that a significant fraction of particles will move 
freely (with energy $ > V_M$). These mechanisms lead naturally to ballistic
motion for $U > U_c$, where the potential $V_i$ is now almost
constant apart fluctuations of order ${\cal O}(1/\sqrt{N})$.

The fact that in the asymptotic limit $(t \to \infty)$ normal diffusion is 
recovered constitutes a typical signature of a noisy dynamics 
\cite{flo1,bet1,kon1}. In order to avoid artifacts due to numerical
noise, in the implementation of the integration 
scheme for the model (\ref{ham}) we took care to maximixe
the numerical precision.  Therefore, the transition
from anomalous to asymptotic ordinary diffusion is attributed 
to a ''deterministic source of noise'', that is intrinsic
of our system and due to finite size effects, 
as we'll show in the next section.

\subsection{Finite $N$ effects}

In this section, we consider finite $N$ effects in order to understand the asymptotic
time dependence of the MSQD outlined in the previous section. As a matter of
fact, for finite $N$, due to the crossing from anomalous to normal diffusion
on long times, a standard diffusion coefficient $D$ can be always defined in 
the limit $t \to \infty$. Therefore we will focus our analysis on the
$N$-dependence of $D$ and of the crossover time $\tau$. 

A deterministic anomalously diffusing dynamical 
systems submitted to weak environmental white noise shows
a transition from anomalous diffusion to standard diffusion on
sufficiently long time scales \cite{flo1}. 
The crossover time $\tau$ for this transition is explicitely computed 
\cite{flo1} and turns out to increase as an inverse power of
the noise amplitude, when short time behaviour is superdiffusive. 
In this context, $D$ is shown to be directly proportional to a  
certain power of $\tau$ \cite{bet1}.

Starting from this knowledge, we can understand the nature of the 
mechanism that generates asymptotic standard diffusion in model (\ref{ham}). 
In order to give an unambigous definition of $\tau$, we consider the local slope
of ${\rm ln}(< r^2(t)>)$ as a function of ${\rm ln}(t)$. The 
crossover time $\tau$ 
is determined when this slope becomes smaller then a threshold value $\mu$.
The diffusion coefficient $D$ can be related to $\tau$, assuming that 
$\tau \propto t_v$ (as already mentioned). By the definition of the 
diffusion coefficient we have $ D \propto \int_0^\infty Z(t) dt $.
Assuming that $t_v$ is sufficiently long, the following
relationship is then straightforwardly found
\begin{equation}
D \propto \tau^{\alpha-1}.
\label{relDtau}
\end{equation}   
This result is in agreement with theoretical predictions
and is sucessfully confirmed by numerical simulations of noisy 
maps \cite{bet1}.

In Fig. \ref{fig8} we report the $N$ dependence of $\tau$ for $\mu = 1.1$
and for two energy values, namely $U=1.48$ and $U=2.0$. In both cases,
we find that $\tau \propto N$.
Moreover, this dependence is not related to the chosen value for the 
threshold $\mu$.  We have indeed verified that for $\mu = 1.2$ no qualitative 
difference could be detected. The interpretation of the $N$ dependence 
of the crossover time is straighforward if we consider finite $N$ effects 
as a source of noise in our model and we can directly deduce that
the amplitude of the noise is typically of order $1/\sqrt{N}$.
This last assumption is justified by the fact that the microscopic
dynamics of the particles generate stochastic fluctuations 
$O(1/\sqrt{N})$ in the values of $M$ and $P$.
These fluctuations become weaker for increasing $N$ and thus naturally 
yield an increasing value of $\tau$. 

The idea of a dynamical noise source due to finite $N$ is also confirmed 
by the analysis of $D$, that turns out also to be an increasing function of
$N$. Some data are reported in Fig. \ref{fig9} for $U=1.48$ and 
$U=2.00$. In particular, considering systems with $ 100 \le N \le 10,000$
we find $D \propto N^\gamma$ with $\gamma = 0.7 \pm 0.1$ and
$1.0 \pm 0.1$ for $U=1.48$ and 2.00, respectively.
These numerical evidences suggest that
\begin{equation}
D \propto {\cal V}^{-\gamma}
\label{variance}
\end{equation}   
where ${\cal V}$ is the variance of the white noise applied on the system.
For model (\ref{ham}), we have ${\cal V} \propto 1/N$. 
Being $\tau$ proportional to $N$, from Eq. (\ref{relDtau})
we expect that 
\begin{equation}
\gamma = \alpha - 1 \quad . 
\label{gamma}
\end{equation}   
However, we find also a dependence of the 
$\alpha$-values on $N$, not negligible at least for $N < 3,000$.
But for $N > 3,000$ a saturation to asymptotic values
is finally achieved (as clearly shown in Fig. \ref{fig10}).
Assuming for $\alpha$ their asymptotic values, we obtain from
the relation (\ref{gamma}) the following $\gamma$-values:
$\simeq 0.64$ for $U=1.48$ and $\simeq 0.9$ for $U=2.00$.
In view of the finite $N$ limitations, we consider
these results as consistent with the numerical results.
A relation analogous to 
(\ref{variance}) was previously found in \cite{vulp} for the
eddy diffusivity associated to a three dimensional noisy 
velocity field \cite{zas2}. Moreover, the authors of Ref.
\cite{vulp} have shown that Eq. (\ref{gamma}) holds
also for their model. Due to the complete different nature
of the two system, we expect that Eqs. (\ref{variance}) 
and (\ref{gamma}) should have some more general field of 
applicability.

In the context of high dimensional Hamiltonian systems, previous results
showed that the crossover time to normal diffusion is inversely proportional to 
the diffusion coefficient when short time behaviour is subdiffusive 
\cite{kan2,kon2}. 
The crossover to standard diffusion is then interpreted as a consequence
of the destruction of the self-similar structure of the stability islands
in phase space.
We believe that the asymptotic normal diffusive behaviour that we observe 
has the same origin as in \cite{kan2}. But let us discuss more in detail
the dynamical mechanisms that are present in our case.

For finite $N$, due to the self-consistent character of model
(\ref{ham}), the single particle potential
fluctuates in time with typical amplitude $O(1/\sqrt{N})$. 
The values of the saddle points thus also fluctuate in time 
and naturally generate a time pulsating separatrix sweeping a phase space domain 
of width $O(1/\sqrt{N})$. Hence, a particle with energy close to $V_s$ can cross 
the separatrix and stochastically experience trapped and channeling motions.
We showed that transport is anomalous in $CP$ and relies on the channeling 
particles that exhibit localized motion when they bounce back 
and forth on the maxima of the potential. But, for finite $N$,
the pulsations of the separatrix induce a second localization mechanism,
that tends to trap the trajectories of IEP into the potential well.

To better illustrate the different dynamical behaviours occuring in the
present system, let us focus on the dynamics of three typical particles, 
namely an initially HEP, IEP and LEP. 
We register for each of them the time evolution of 
(a) their orbit in the $(x,y)$ plane, 
(b) their coordinate $x(t)$ and (c) the single particle energy $h_i$ 
as defined in (\ref{ham1}). 
The results are shown in Fig. \ref{figlevyener} where the first column
corresponds to HEP, the second to IEP and the third to LEP. For each
particle (a), (b) and (c) are plotted in the first, 
second and third row, respectively.
Visual inspection of HEP(a) and IEP(a) shows a quiet 
similar behaviour and an enlargement of these trajectories 
would yield a picture similar to the one already reported in
the inset of Fig. \ref{fig5}(a). However, Fig. 
\ref{figlevyener}-IEP(b) indicates that 
the initially IEP experiences long time localized sequences
that are not present for the HEP.
Fig. \ref{figlevyener}-IEP(c), showing the time
dependence of the energy of IEP, clearly indicates that in the long lived
localized regime corresponding to $t \in [10000, 17000]$, the particle 
is trapped since its energy $h_i$ goes below the average energy of the 
separatrix  $<V_s>$ (indicated in this figure by a full line).

The localization within the potential well, due to separatrix crossing,
interrupt the sequence of correlated flights and localizations along the 
channels, and naturally inhibits superdiffusion. This inhibiting 
effect of separatrix crossing is due to finite $N$ and its consequences 
on the numerical value of the exponent $\alpha$ is illustrated in 
Fig. \ref{fig10} for two values of $U$. For given $U$, this figure shows an increase 
of the order of $10 \%$ in $\alpha$ when raising $N$ from $10^3$ to $10^4$.
The exponent $\alpha$ then saturates to an almost constant value indicating
that the effect of trapping and untrapping by the fluctuating
separatrix decreases and finally becomes negligible for growing $N$.
In other words, as $N$ grows, 
the phase space volume swept by the pulsating separatrix shrinks and finally
vanishes in the mean-field limit $N \to \infty$. 
As a final remark, we should also notice that the channeling
particles can be "decorrelated" not only by the trapping mechanism
but also by an escaping mechanism to higher energies (for $h_i > V_M$). 
But from numerical evidence we can conclude that this other mechanism
is less relevant, at least on the time scale of our simulations.

Finite $N$-effects are responsible, besides of the fluctuations of the
single particle potential, also of a second mechanism 
generating "dynamical noise".  In order to explain it more in detail 
let us consider the LEP particle dynamics:
a typical LEP-orbit is reported in Fig. \ref{figlevyener}.
One observes essentially two main aspects of its dynamics: 
(i) the extremely slow
motion of LEP with respect to IEP and HEP (cf. the scales); 
(ii) a drifting
motion of LEP that resembles a brownian motion.
During all the simulation time  ($t = 30,000$)
the corresponding energy, plotted 
in Fig. \ref{figlevyener}-LEP(c), remains close to the average 
value of the potential minimum $<V_m> \approx 0.654$.
This indicates that the particle remains trapped during 
the entire simulation. Being for $U = 1.0$ the crossover time
$10^4 < \tau < 10^5$, the previous result suggest that such a trapped
particle will eventually escape from the well on time scales
much longer than $\tau$.
We moreover verified that this observation remains true for almost 
any particle 
with initial energy smaller that $0.5 V_s \approx 1.85$. Which  
for $U=1.00$ represent approximatively $75 \%$ of all the particles. 

The above observations confirm the fact that anomalous
diffusion is only due to the small fraction of particles that
evolve inside the channels of the potential.
The trapped particles, indeed, due to their
extremely slow motion, contribute negligibly to the average MSQD.
But the slow drift of the clustered particles,
corresponding to a drift of the potential seen by a single
particle, inhibits flights on very long times. This because a
channeling particle will not have a free horizon in front of itself
at any time, as it happens for a particle moving in a fixed
potential frame.

The origin of the collective drifting motion of the clustered 
particles can be understood in the following way:
let us assume that at time $t=0$ all the particles are trapped
in the potential well and that the total momentum is zero and
that at a later time $t>0$ one particle escapes from the cluster.
This particle will carry out of the cluster a non-vanishing average 
momentum $p$, but as the total momentum should be conserved, 
the cluster will then start to move with an average momentum $-p/(N-1)$.
Therefore, in the mean-field limit this effect will disappear and
consistently we will have anomalous diffusion at any time \cite{note}.

The two mechanism outlined above, will have a cumulative effect
on the phase space topology of our system.
Due to the analogy with the egg-crate potential, in the mean-field limit 
the phase space of our model will exhibit a hierarchy of
nested self-similar stability islands \cite{geis2,geis3,kla1,fle1,cha1}.
The monotonous increase of the value of $\alpha$ with $N$ reported in
Fig. \ref{fig10} intuitively suggests a similar continous picture 
when considering
the modifications occuring in the phase space topology under
continous variation of $N$. In particular, for finite $N$ we expect 
that the phase space structures associated to the smallest islands 
disappear up 
to a typical size that decreases for growing $N$.
Being the self-similarity no more
complete in phase space, normal diffusion should be recovered
beyond a crossover time that grows for increasing $N$.

\section{Lyapunov analysis}

In order to complete the description of model (\ref{ham}), 
we investigate in this section 
a fundamental indicator to characterizes the dynamics 
of Hamiltonian models: the maximal Lyapunov exponent $\lambda$.
Our analsys rely on numerical estimation of $\lambda$,
performed considering the evolution in the tangent space
of the model and applying a standard technique 
introduced in Ref. \cite{benettin}.

Our model is integrable in the limit
of low and high energy, therefore $\lambda \to 0$
for $U \to 0$ and $U \to \infty$. In between these
two extrema we expect that a finite Lyapunov exponent
will be observed similarly to what was recently found for
1D self-gravitating models \cite{yama,rapi1,firpo}.
Our data are reported in Fig. \ref{fig11} for three
different types of initial conditions :
\begin{itemize}

\item
(A) the particles are initially clustered and the velocity
distribution is Maxwellian;

\item
(B) the velocity distribution is again Maxwellian 
but with a thermal velocity conciding with its canonical
prediction and the particles are organized in a single cluste in 
such a way that also the $M$- and $P$-values coincide with their 
canonical prediction;

\item
(C) the particles are initially clustered with a water-bag
velocity distribution.

\end{itemize}

The initial condition (C) is the one commonly  used through the present
paper, in particular for the study of transport in the system.
As can be seen fom Fig. \ref{fig11} $\lambda$ grows for
increasing $U$ up to a maximum value and then 
decreases. Such maximum (at least for $N=200$) is located
at an energy $U \simeq 1.3-1.4 < U_c$.
For $U > U_c$, we observe a power law decrease of 
$\lambda$ with $N$. In particular, for $U=3.0$
we found a power law exponent $\sim 0.31$ (see Fig.
\ref{fig12}), in good agreement with recent theoretical results 
obtained via a random matrices approach in Ref. \cite{rapi1} 
and with a Riemanian geometrical technique \cite{firpo}. In both
these studies an exponent $1/3$ has been found 
considering a model similar to (\ref{ham}) in 1D \cite{ruf}. 

In the low energy limit (for $U < 0.01$) a power
law increase of the type $U^{1/2}$ is clearly 
observable for all the three types of initial conditions.
The data are reported in the inset of Fig. \ref{fig11}.
A similar behaviour was found for a 1D
mean-field model \cite{rapi1}. This indicates
that for fully coupled Hamiltonian systems this
property holds in general, independently of the
space dimensionality. In particular, for $U \to 0$
the particles are all clustered, therefore we expect
that the scaling $\lambda \propto U^{1/2}$ should
be related to a "collective" chaotic mechanism.
Work is in progress in order 
to derive a theoretical explanation of such behaviour
\cite{ruf2}.

Let us now try to understand which are the mechanisms
underlying the observed behaviours of $\lambda$.
Once the energy
$U$ is fixed for all the three considered initial 
conditions, after a reasonable transient, we obtain
exactly the same value for the average temperature 
$T$ and magnetizations $M$ and $P$. However, 
at low energies ($U < 0.8$) the measured
average $\lambda$ depend heavily on the initial
conditions. This clearly indicates the coexistence
of several equivalent state, that can be considered
as equilibrated within the examined time interval.
Usually we have averaged the maximal
Lyapunov over a time $1,000,000 < t < 9,000,000$
after an equilibration time ranging from $t = 500,000$
to $t = 20,000,000$ (this last value has been used
in particular for extremely low energies).
Obviously, also if the considered time scale are
considerably long we cannot exclude that these
states are metastable. It should be noticed that this
kind of behaviour is unexpected in $N$-body Hamiltonian
systems, because it is commonly believed that for sufficiently
high $N$ Arnold diffusion takes place and each orbit
is allowed to visit the complete phase space. But our
data instead indicate that some "barrier" in the
phase space still survive even for $N =200$. 
The origin of this lack of ergodicity is related
to the long range nature of the forces that
induces a persistent memory of the initial
conditions, as previously noticed by Prigogine 
and Severne \cite{prigo} for gravitational plasmas.
Recently, some numerical evidence of non ergodicity
has been reported also for a 1D mass sheet model
\cite{gouda}.

It is clear from Fig. \ref{fig11} that, in the
interval $U \in [0,0.8]$, for initial conditions 
of type (B) $\lambda$ remains always smaller
than the corresponding exponents obtained with
initial conditions (A) and (C).
The maximal differences
are observed in the energy range $0.2 < U < 0.8$,
where particles begin to escape from the cluster
(this for initial conditions (A) and (C)).
Above $U \simeq 0.9$ the same Lyapunov is obtained
for all type of initial conditions. A typical feature
of the initial conditions (B), for $U < 0.8$ is that
all the particles are trapped in the potential well.
Instead when
one or more particle escape from the cluster ($U > 0.9$)
also with this initial condition the usual $\lambda$ is
obtained. We believe that two chaotic mechanisms are present
in the system: one felt from the particles moving
in the minimum of the potential and one from particles
visiting a region near to the separatrix. This second
mechanism is known and is related to a chaotic
belt present around the separatrix \cite{ll}.

The first one should be instead due to the temporal
erratic oscillations of the minimum of the potential well.
In order to clearly identify such mechanisms we have followed
the trajectory of a system initially prepared with condition
(B) for an energy $U = 0.87$. On short time all the particles
are trapped and we measure an average value $\lambda \simeq 0.13$. 
At a later time one particle escapes from the cluster 
and $\lambda$ shows a jump to a value that is almost double 
(see Fig. \ref{fig13}). 
In Fig. \ref{fig13} the magnetization $M$ and the kinetic 
energy $K$ are also reported. When the particle escapes
from the cluster $M$ shows a clear decrease as well as $K$.
This last effect is due to the fact that the potential energy $V$
is minimal when all the particles are trapped, therefore if
one escapes $V$ increases and due to energy conservation
$K$ decreases. This is the phenomenon at the basis of the
negative specific heat effect. From the above arguments
we can identify a strong chaos felt from the particles
approaching the separatrix and a minimal chaos associated
to orbits trapped in the minimum of the potential.
The presence of these 2 chaotic mechanisms together
with the non-ergodicity of the system explains the
strong dependence of the $\lambda$-values
from the starting conditions.

As a final point,
we would like to notice that in the CP 
for low energy density the Lyapunov exponent 
averaged over short times
exhibits an intermittent behaviour, when 
starting conditions (A) or (C) are considered.
In particular in Fig. \ref{fig14} the instantaneous 
$\lambda$ is shown for condition of type (C) together
with its running averaged value and the corresponding
averaged value for condition (B). 
The intermittent behaviour can be explained as due to 
trapping and untrapping of the IEP's.
As a matter of fact, for initial conditions (B) and
energies $U < 0.8$ no diffusion
at all is observed (the MSQD saturates to a constant
value for long times), while for initial condition (C)
(or (A)) a super-diffusive motion is observed for 
energies higher than $U=0.3$. This confirms that
channeling particles (i.e. IEPs) are affected by a stronger 
chaotic mechanism than the trapped ones (namely, the LEPs).

\section{Conclusions}

In the present article we have analyzed the equilibrium
and dynamical properties of a 2D $N$-body self-gravitating Hamiltonian 
system.
Our main result is the occurence of a thermodynamical phase transition 
associated to a dynamical transition from anomalous to ballistic transport.

Firstly, the statistical equilibrium description in the canonical ensemble
has been reported. Revealing a first order phase transition between a collapsed 
phase, characterized by the presence of a single cluster of particles, and 
an homogeneous phase where particles are uniformly distributed. Close to
the transition, the presence of two equivalent minima in the Helmholtz free 
energy clearly indicate a regime where the two phases (CP and HP)
coexist. Within the microcanonical ensemble, for energies close to $U_c$ 
MD results reveal a negative specific heat regime. Referring to
previous work \cite{her1}, we interpret this phenomenon as a signature of 
non-equivalence between canonical and microcanonical statistical ensembles.

Secondly, we have shown that in a broad interval of energies 
$0.3 < U < U_c$ the transport in this system is anomalous. In the 
sense that the particles show a super-diffusive motion below a
cross-over time and normal diffusion for longer times. Usually,
normal diffusion is expected to occur in $N$-body systems due to
the absence of long time correlations. Instead in the present
case the non markovian nature of the process is clearly evidenced
by the occurence of Levy walks and of power law decay for the
VACF. In the CP, from simple mean-field considerations one can conclude
that each particle evolves in a $2-$D time dependent egg-crate 
like potential. Channeling particles are at the origin of anomalous 
diffusion on short time scales. Since their motion 
is made up of flights along the channels interrupted by localizations
due to bouncing of the particles on the maxima of the potential.
We have also shown that finite $N$ 
effects generate a pulsating separatrix and give rise to a second 
localization mechanism for the particle motion
due to the trapping of the orbits in the potential well.
This phenomenon together with the slow drift of the potential well
is responsible for the suppression of anomalous diffusion on long time scales.

These two finite $N$-effects can be interpreted as a white noise source
affecting the single particle dynamics. As a matter of fact these effects
become weaker for increasing $N$, as confirmed by the linear dependendence
of the crossover time on $N$. The asymptotic dynamics of the model then depends 
on the order the two limits $N \to \infty$ and $t \to \infty$ are taken. 
If we first take the limit $N \to \infty$ and then $t \to \infty$ diffusion remains 
anomalous for all time. Ohterwise standard diffusion is recovered for 
sufficiently long time.

The Lyapunov analysis shows clearly that even for $N=200$ the phase space
is not fully accessible and the visited portion of the phase space
depends on the initial conditions. This is quite unexpected in high
dimensional Hamiltonian systems.  Moreover, two different chaotic mechanisms
have been identified:
a minimal chaos associated to the clustered particles and a strong chaos
affecting the IEP particles. Anomalously diffusing particles have therefore
an intermittent chaotic behaviour. 

We believe that the properties displayed by the model (\ref{ham}) 
are quite general for gravitational systems. It is also
our claim that the inclusion of higher order Fourier harmonics in 
the potential will not qualitatively affect the results here presented. 
This is indeed confirmed by a recent study by Abdalla and 
Reza Rahimi Tabar \cite{abdulla} for the 2D full logarithmic 
Newtonian potential $V_\infty$. Also in that case a transition 
at finite $T$ from a clustered to a homogeneous phase is 
still present.
This constitutes in our opinion an interesting basis for the 
generalization of the present results to models that mimic more 
realistically the gravitational interactions (e.g. systems with true 
Newtonian potentials in 3D).

Due to the strict analogy of the model here studied with point
vortices model for 2D turbulence \cite{fri1} we expect that
anomalous diffusion should also be observable for such 
models. Preliminary indications \cite{prov1} seem to confirm 
our claim and ask for more accurate investigations in such
direction.

\acknowledgments
We would like to thank L. Casetti, P. Cipriani, A. Rapisarda, 
P. Poggi, H.A. Posch, R. Trasarti-Battistoni and 
A. Vulpiani for useful discussions 
and positive interactions. In particular we are indebted with 
S. Ruffo and M.-C. Firpo for a careful reading of the paper.
Institute for Scientific Interchange (I.S.I) in Torino (Italy)
is also acknowledged for the hospitality offered to A.T.
during the workshop "Complexity and Chaos", where part
of this work has been completed.  M.A. also thanks 
D. Noack for the important logistical support she provided 
him during his stay in Dresden.

\begin{figure} [h]
\psfig{figure=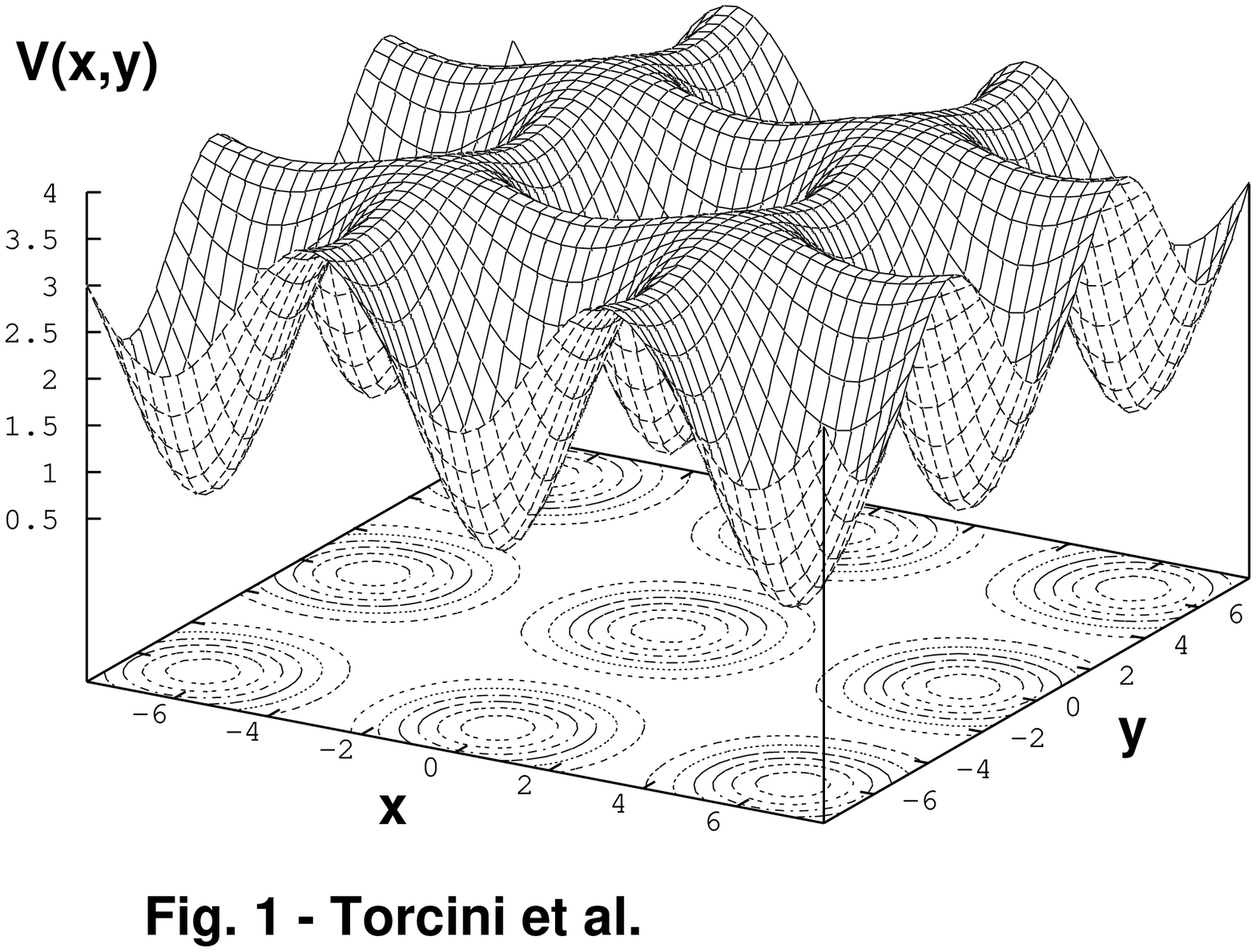,angle=-90,height=7truecm,width=7truecm}
\vskip 0.5truecm
\caption{Instantaneous single particle potential $V(x,y)$ together
with its contour-plot for $U=1.00$ and $(x,y) \in [-5 \pi/2, 5 \pi/2]^2$.
$M_1 = M_2 =0.8249$, $P_1 = P_2 =0.6929$, $\phi_z=0=\psi_z$}
\label{fig1}
\end{figure}

\begin{figure} [h]
\psfig{figure=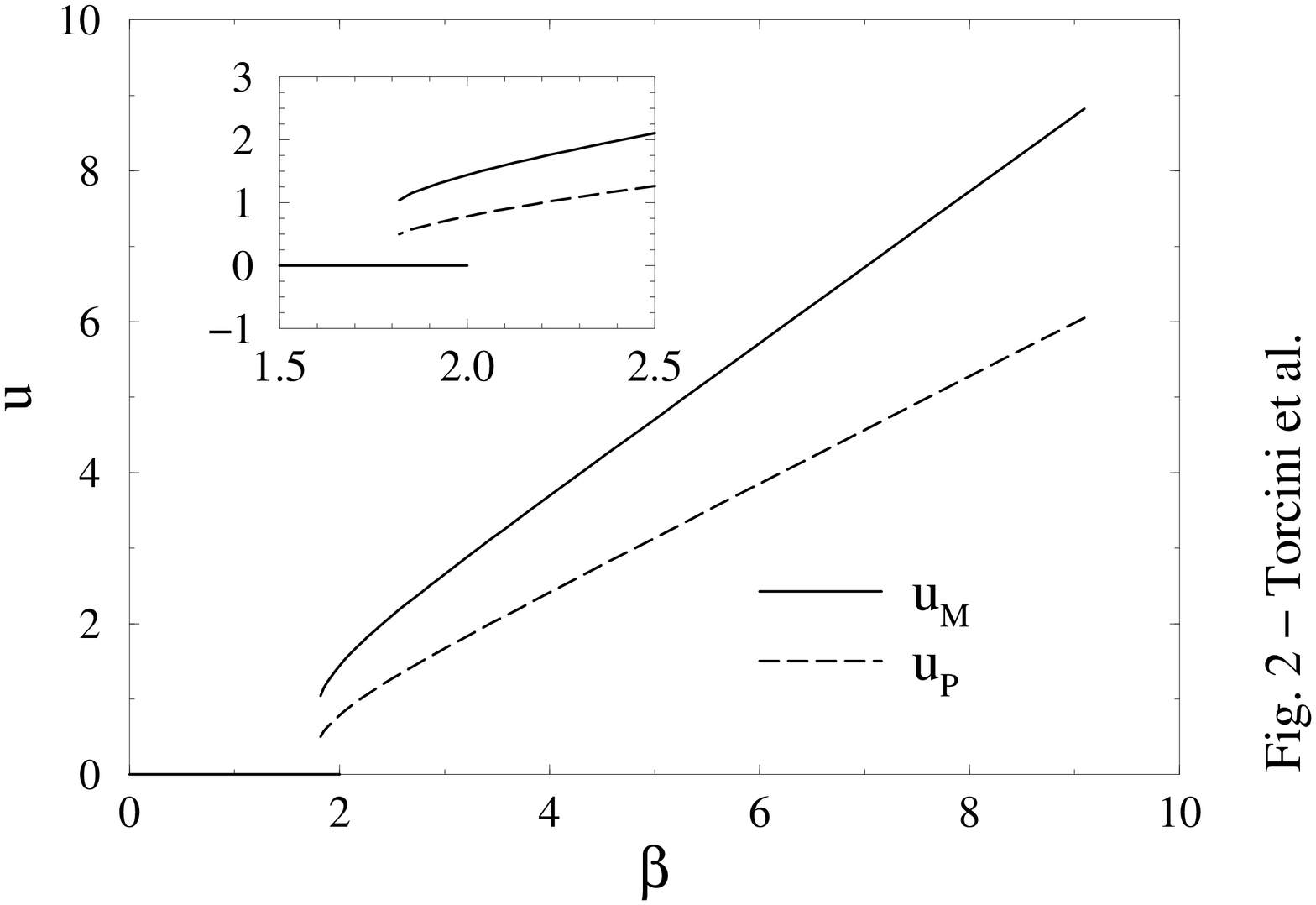,angle=-90,height=7truecm,width=7truecm}
\vskip 0.5truecm
\caption{$\beta$ dependence of ${\bar u}_M$ (full line) and ${\bar u}_P$ 
(dashed line). For $\beta \in [1.81,2]$ there are two coexisting minima: 
one at the origin (${\bar u}_M,{\bar u}_P)=(0,0)$ and a second one for which
$0< {\bar u}_P < {\bar u}_M$. 
}   
\label{fig2}
\end{figure}

\begin{figure} [h]
\psfig{figure=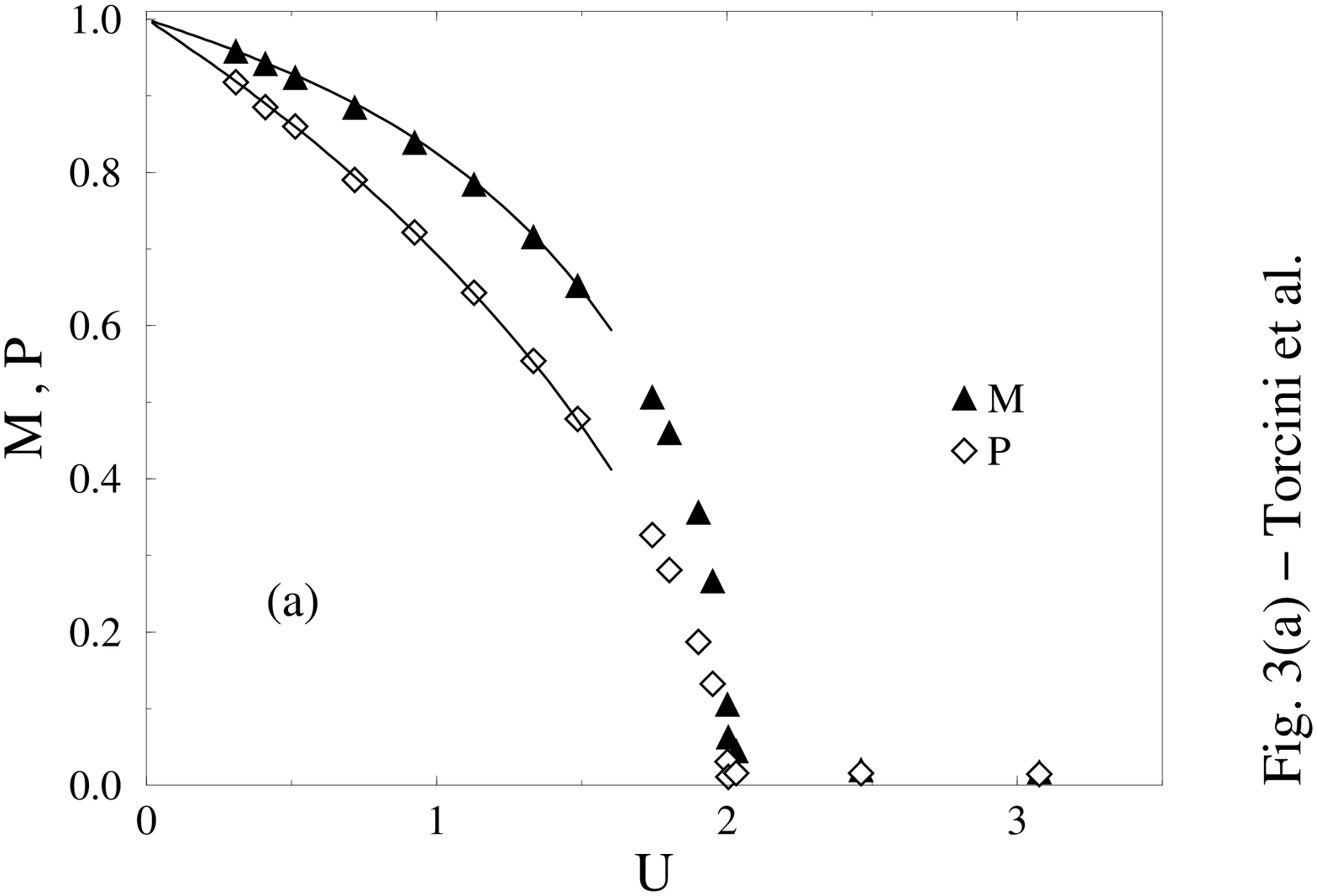,angle=-90,height=7truecm,width=7truecm}
\vskip 0.5truecm
\psfig{figure=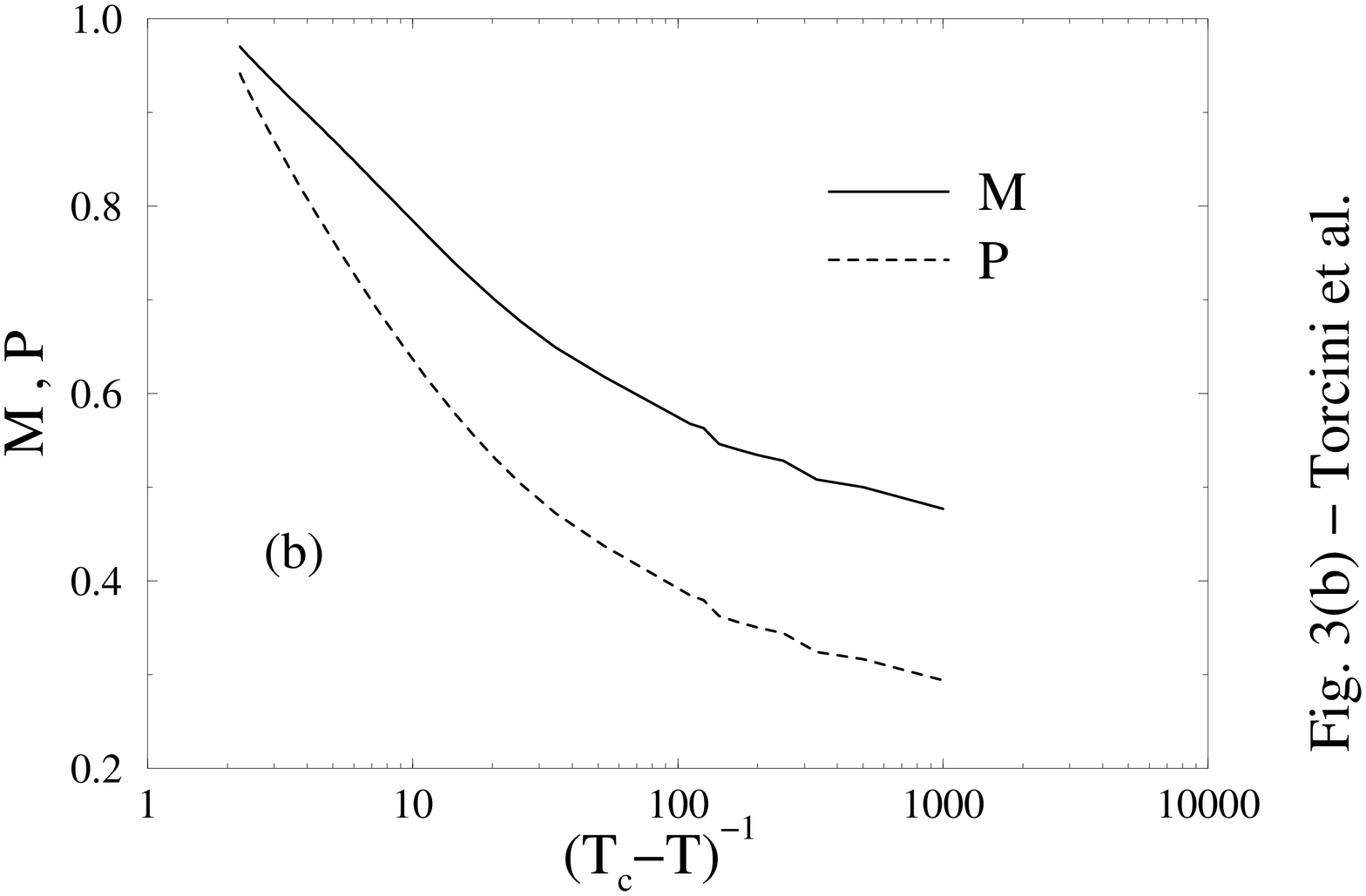,angle=-90,height=7truecm,width=7truecm}
\vskip 0.5truecm
\caption{
In (a), the canonical estimation 
of $M$ and $P$ (solid curve) together with their time averaged values
obtained from microcanonical simulations (symbols) are shown. 
The measurements have been obtained with $N=4,000$ (a part a few point 
with $N=10^4$) and averaged over a total integration time ranging 
$t=1.2 \times 10^6$ to $t=2.4 \times 10^6$ with a time step $dt = 0.3$.   
In (b) are reported the canonical estimation for $M$ and $P$ 
as a function of $(T_c - T)^{-1}$ in a log-linear plot.
Neither $M$ nor $P$ have usual critical exponent.}
\label{fig3}
\end{figure}

\begin{figure}
\psfig{figure=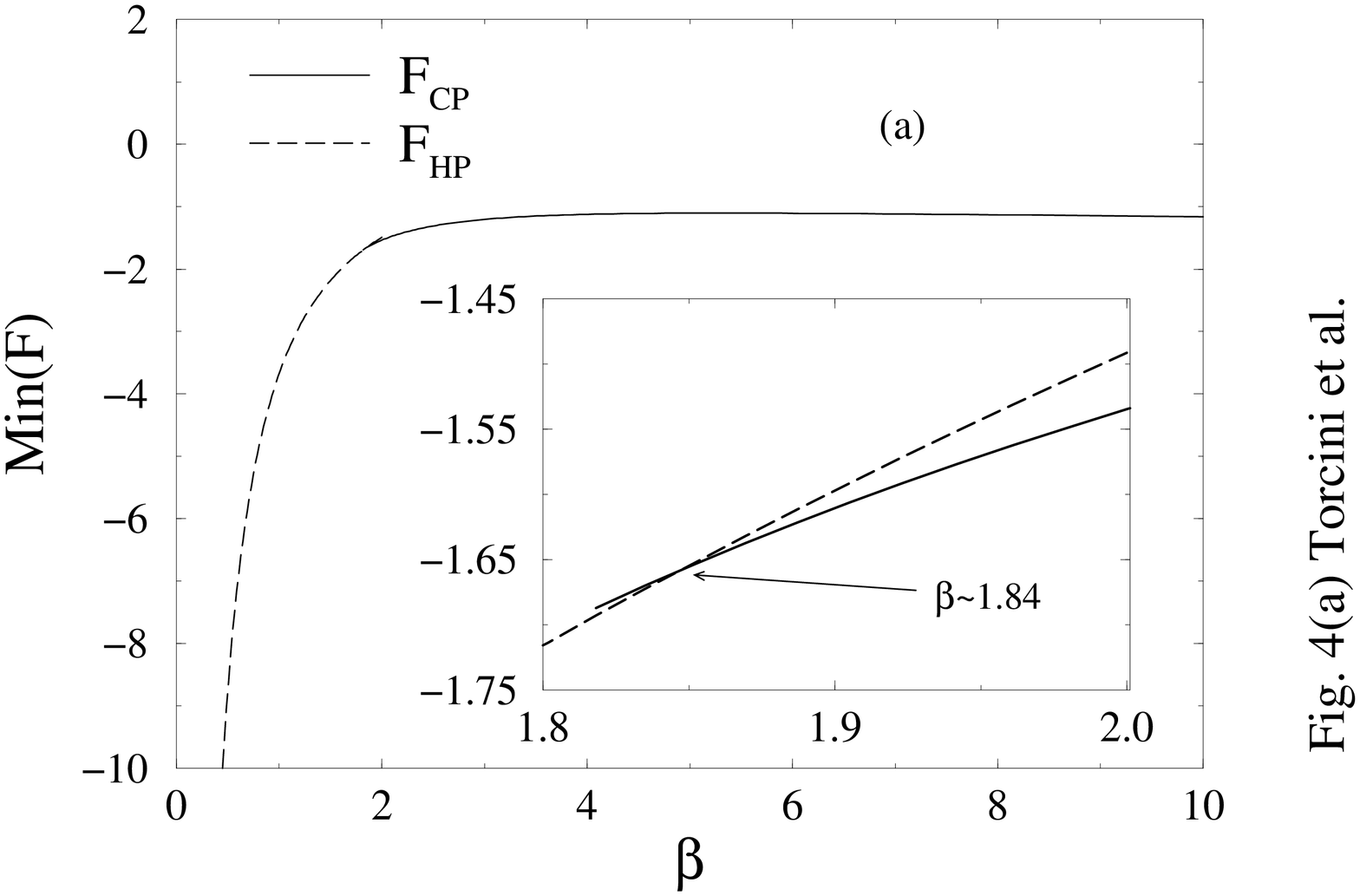,angle=-90,height=7truecm,width=7truecm}
\vskip 0.5truecm
\psfig{figure=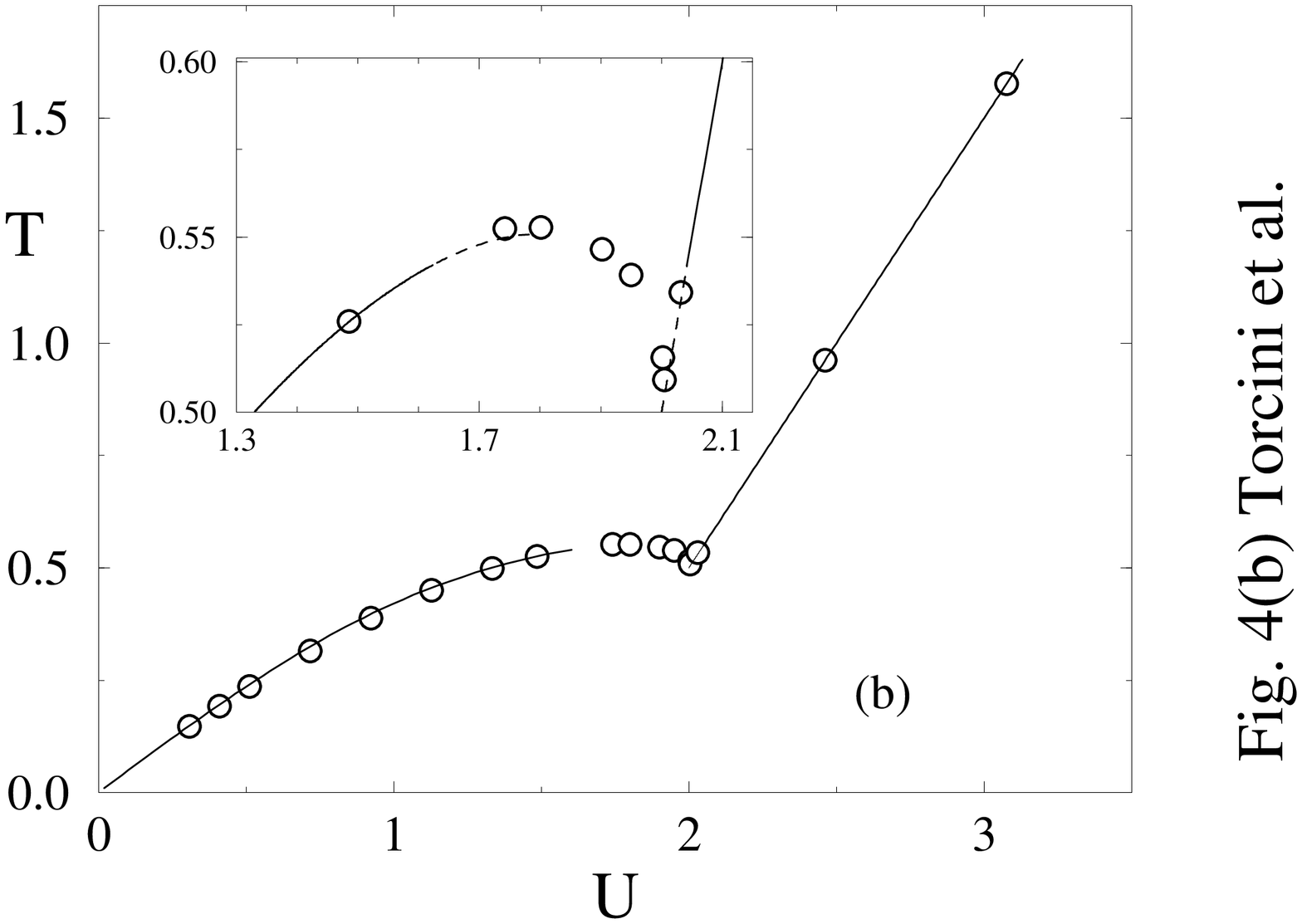,angle=-90,height=7truecm,width=7truecm}
\vskip 0.5truecm
\caption{
In (a) we plot the minimum
value of the free energy function $F$ as a function of $\beta$. 
In the coexisting range 
$F$ has two minima which take the same value at critical inverse
temperature $\beta_c \approx 1.84$ (see inset).
(b) Temperature as a function of the specific energy $U$.
The theoretical predictions obtained within the canonical ensemble 
are indicated by lines. The circles correspond to microcanonical 
numerical results.  The inset is an enlargement of the transition region:
the full curves (resp. dashed) refer to the absolute (resp. relative) 
minimum of $F(T)$. 
} 
\label{fig4}
\end{figure}

\begin{figure} [h]
\psfig{figure=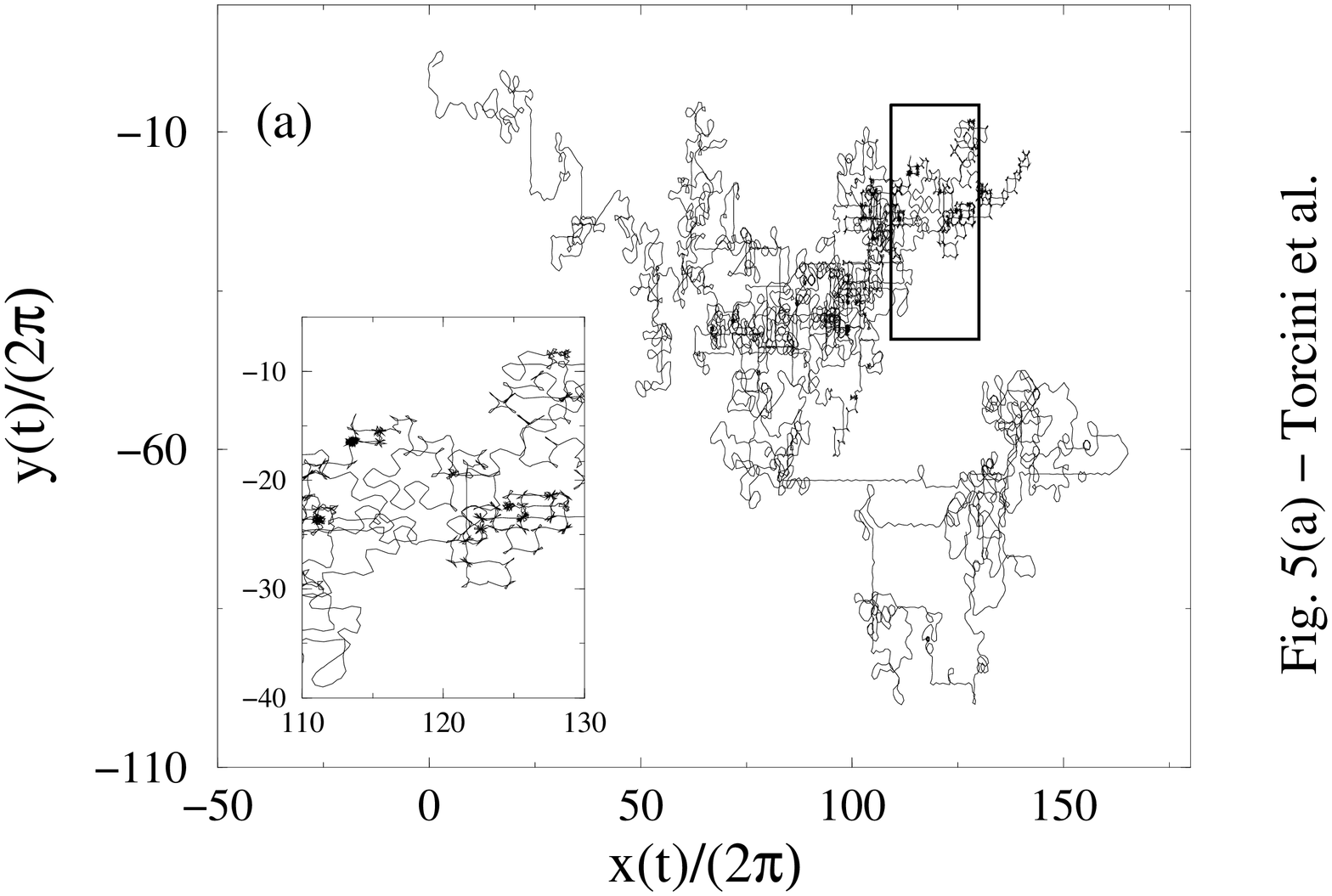,angle=-90,height=7truecm,width=7truecm}
\vskip 0.5truecm
\psfig{figure=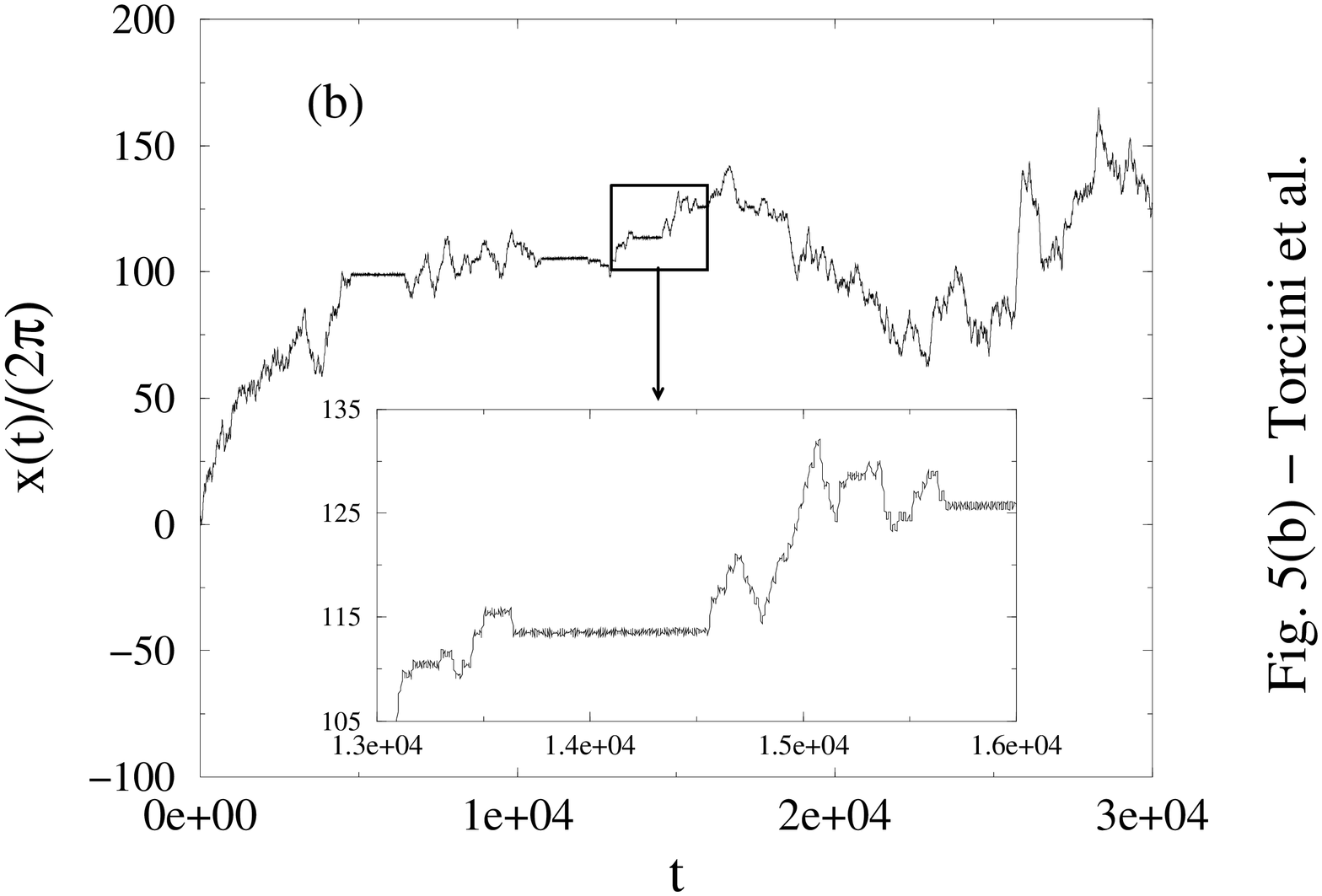,angle=-90,height=7truecm,width=7truecm}
\vskip 0.5truecm
\caption{Typical orbit of an initially IEP in the $(x,y)$ plane (a) for 
$U=1.00$ and $N=4,000$. 
In this representation, the $2-D$ torus on which the dynamics takes place is
unfolded and represented as an infinite plane constituted of an infinite 
number of juxtaposed elementary periodic cell of size $1 \times 1$. 
In the inset an enlargement of the trajectory in the indicated box is
reported.  In (b) we plot the corresponding time 
evolution of coordinate $x(t)$ and the enlargement corresponding to 
the sequence in the inset of (a).}
\label{fig5}
\end{figure}

\begin{figure} [h]
\psfig{figure=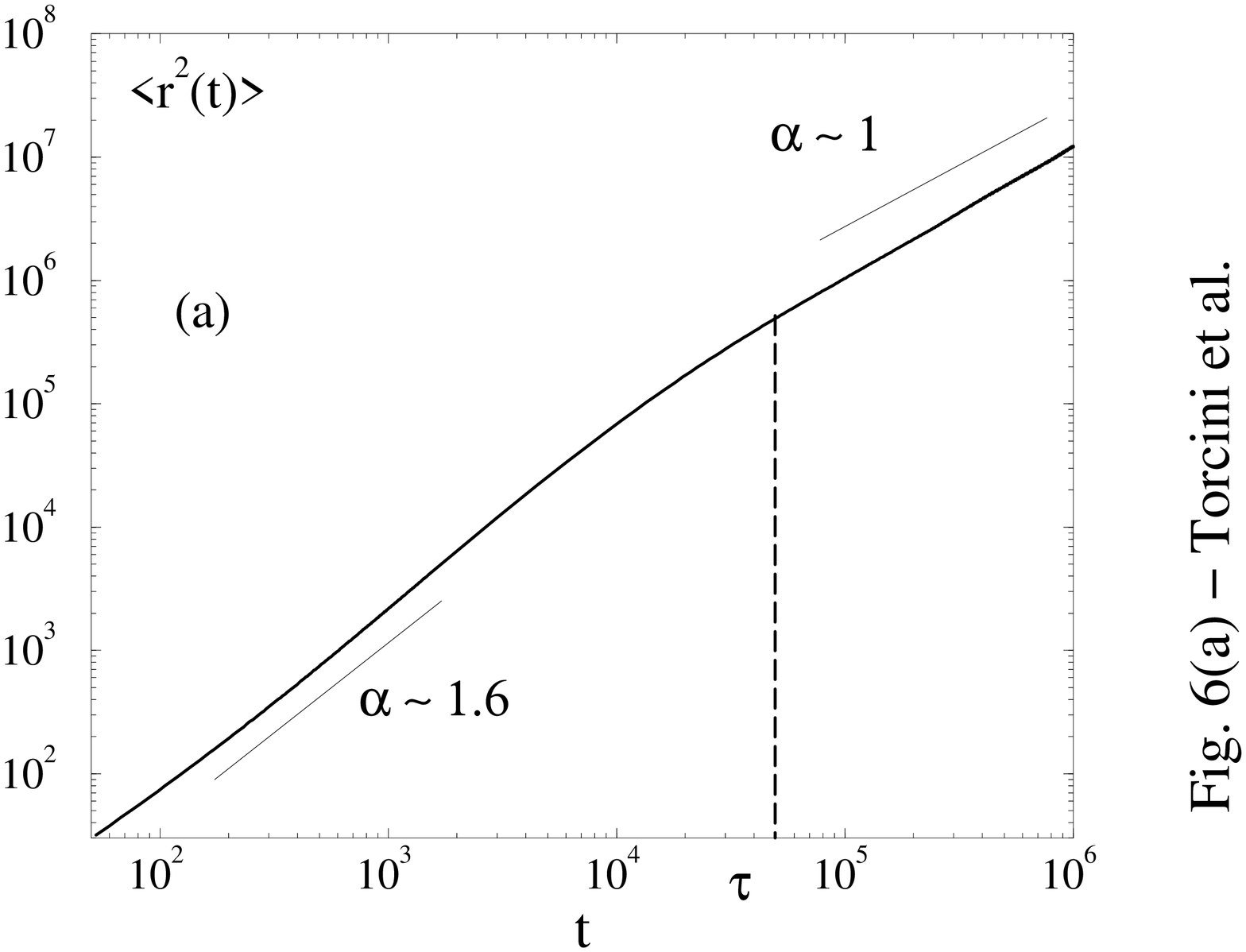,angle=-90,height=7truecm,width=7truecm}
\vskip 0.5truecm
\psfig{figure=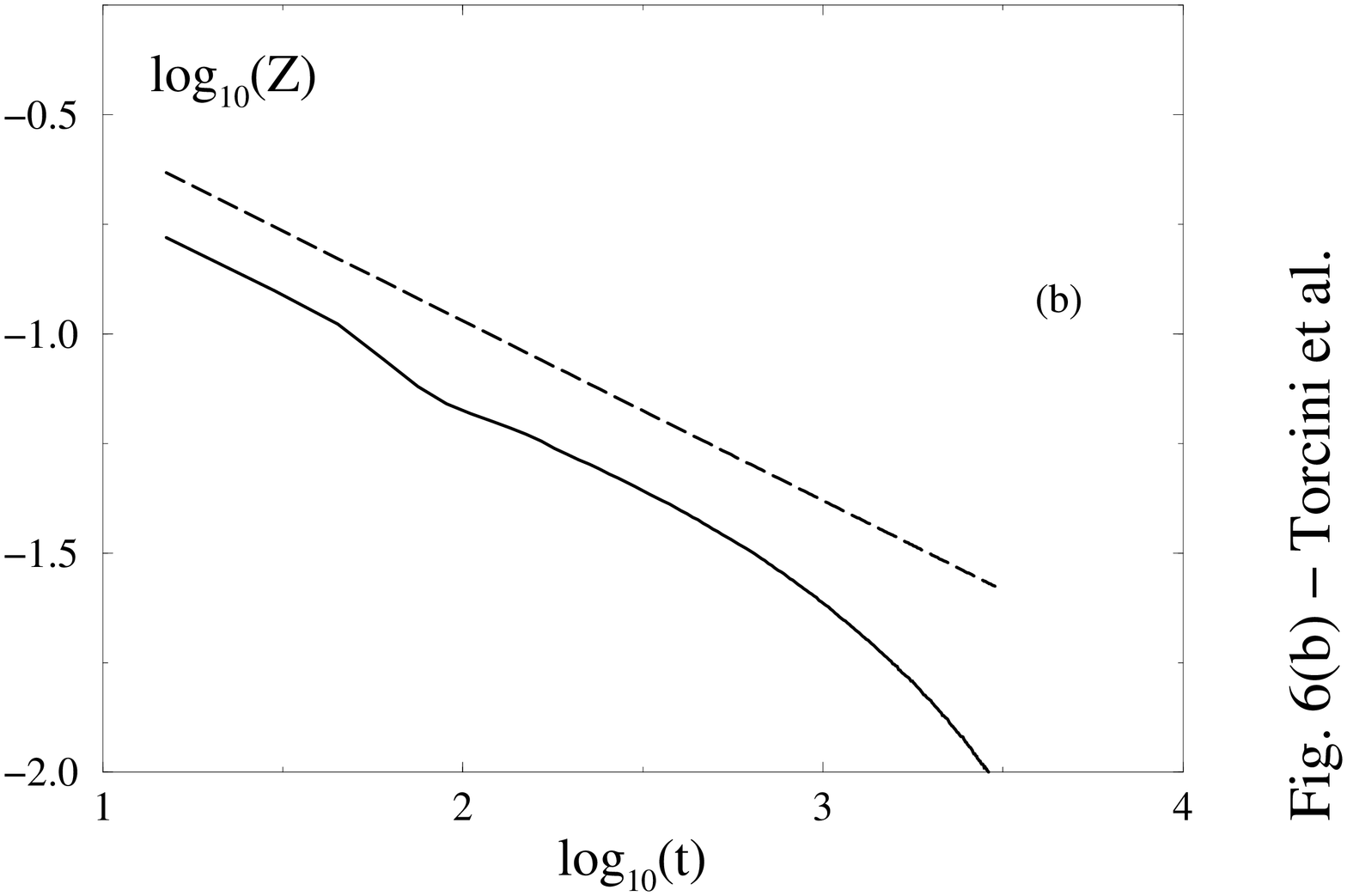,angle=-90,height=7truecm,width=7truecm}
\vskip 0.5truecm
\caption{Time dependence of the mean square displacement $<r^2(t)>$ in a
log-log representation for $U=1.1$ and $N=4,000$. The numerical results
correspond to the solid line. The segments are the estimated slopes 
of $<r^2(t)>$ when $t < \tau$ and $t > \tau$. 
In (b) the logarithm of the VACF is displayed 
as a function of $\log(t)$ for $U=1.74$, in such a
case the estimated $\alpha$-value is 1.59. 
The reported slope is $2 - \alpha =$ 0.41} 
\label{fig6}
\end{figure}

\begin{figure} [h]
\psfig{figure=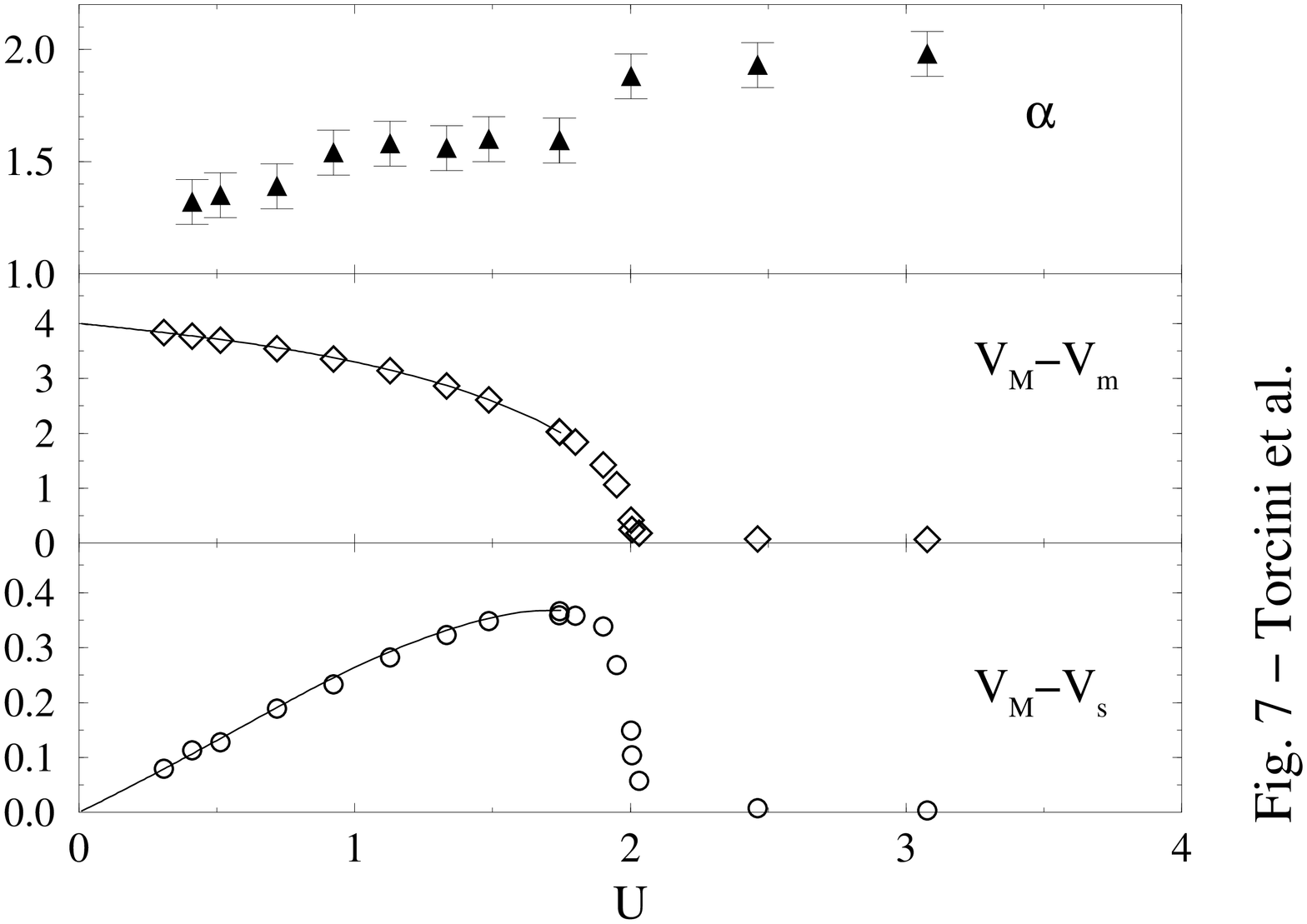,angle=-90,height=7truecm,width=7truecm}
\vskip 0.5truecm
\caption{Coefficient $\alpha$ (triangles), time averaged depth 
of the potential well $V_M - V_m$ (diamonds) and 
of the channels $V_M-V_s$ (circles) 
of the single particle potential 
as a function of $U$. The full line corresponds
to the canonical prediction.
The measurements of $\alpha$ have been obtained with $N=4,000$ 
(a part a few point with $N=10^4$)
and averaged over a total integration time ranging $t=1.2 \times 10^6$
to $t=2.4 \times 10^6$ with a time step $dt = 0.3$.} 
\label{fig7}
\end{figure}

\begin{figure} [h]
\psfig{figure=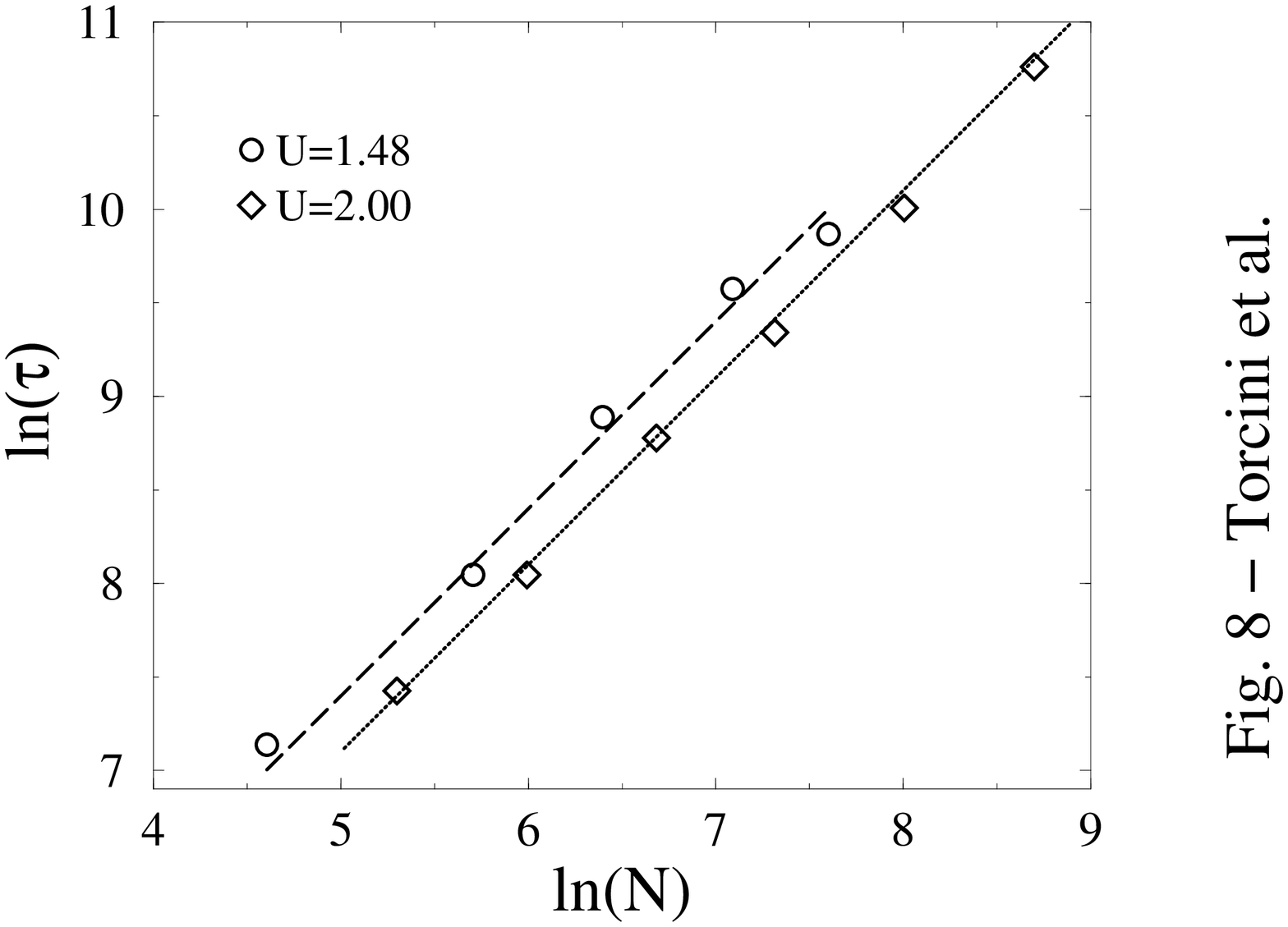,angle=-90,height=7truecm,width=7truecm}
\vskip 0.5truecm
\caption{Crossover time $\tau$ as a function of $N$ for $U=1.48$ 
(circles) and $U=2$ (diamonds). The values of $\tau$ have been estimated 
when the local logarithmic slope of the MSQD decreases
below a threshold value $\mu = 1.1$. The solid line is the best linear
fit to the data. Its slope is $0.95 \pm 0.08$ 
(resp. $0.96 \pm 0.06$) for $U=1.48$ (resp. $U=2.00$).} 
\label{fig8}
\end{figure}

\vglue0.5cm
\begin{figure} [h]
\psfig{figure=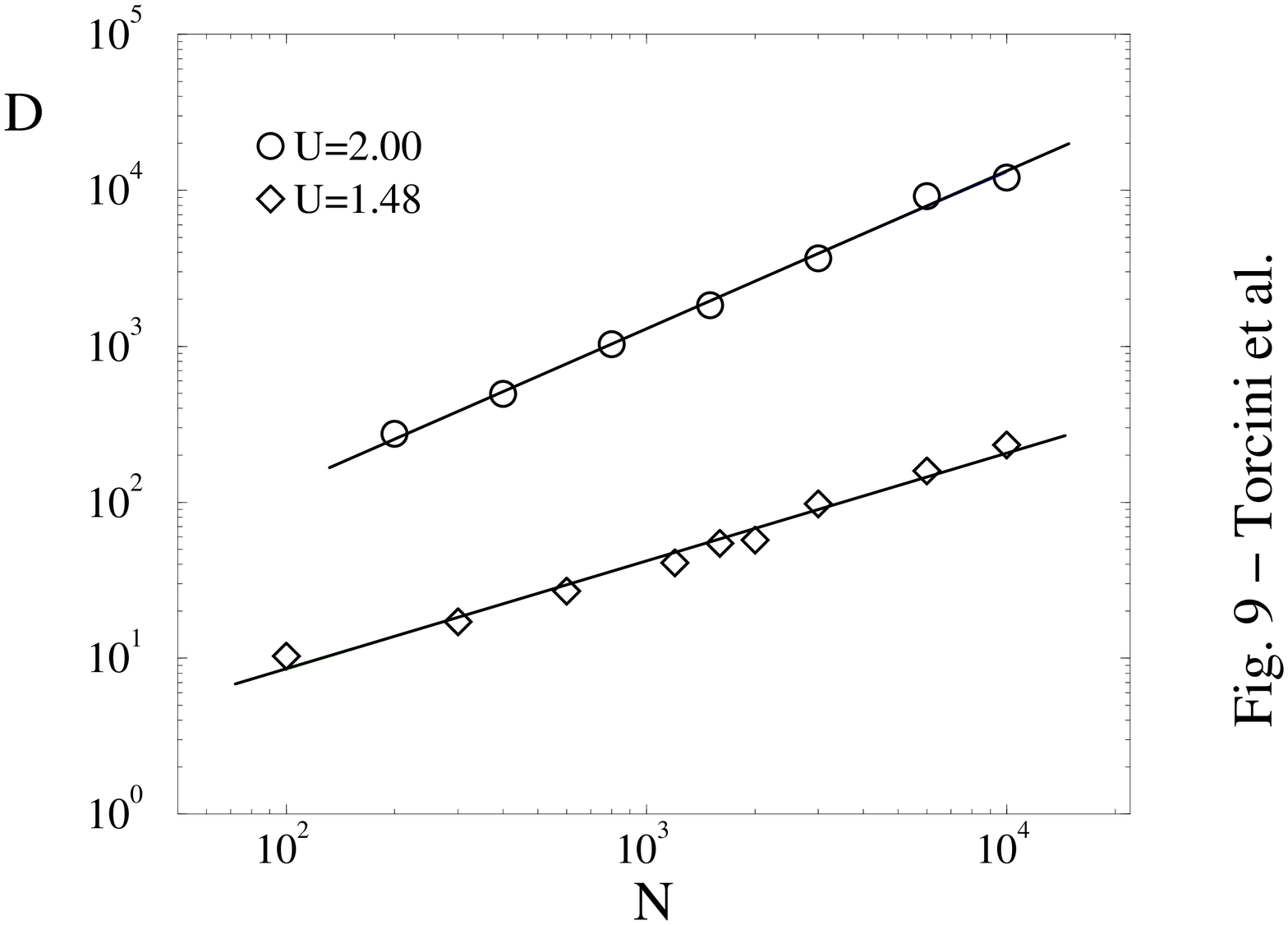,angle=-90,height=7truecm,width=7truecm}
\vskip 0.5truecm
\caption{Logarithm of the diffusion coefficient $D$ as a function of 
$\log(N)$ for $U = 1.48$ (diamonds) and $U = 2.00$
(circles). Straight lines correspond to linear fitting to
the numerical data, the slope are $0.7 \pm 0.1$ 
and $1.0 \pm 0.1$ for $U=1.48$ and $U=2.00$, respectively.
} 
\label{fig9}
\end{figure}

\begin{figure} [h]
\psfig{figure=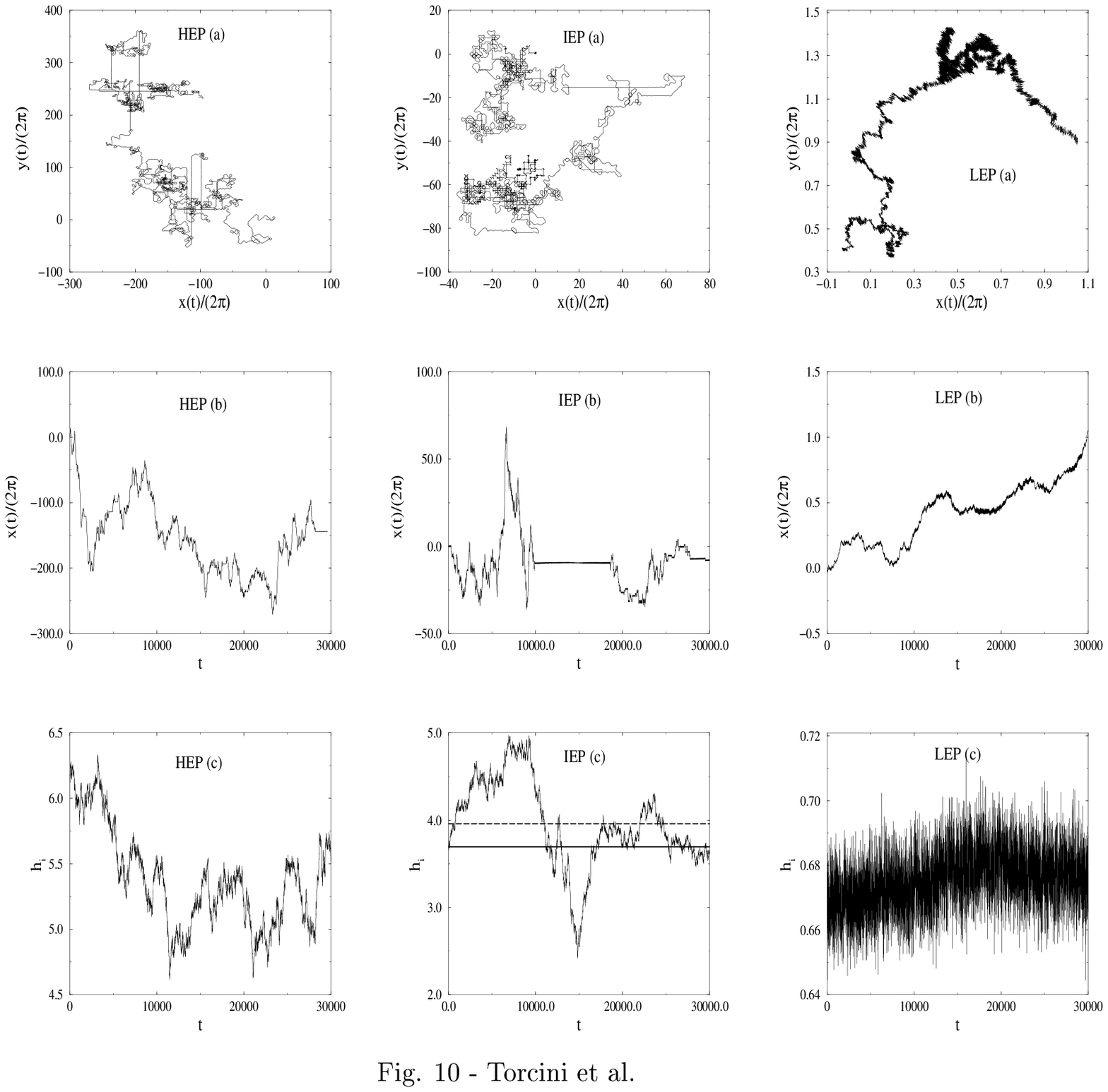,angle=0,height=11truecm,width=8truecm}
\vskip 0.5truecm
\caption{Typical orbits of HEP, IEP and LEP (first row)
together with the time dependence of coordinate $x(t)$ (second row)
and the time dependent energy $h_i$ (third row) of each orbit.
The first column refers to HEP, the second to IEP and the
third to LEP. Note the difference of scale in graphs of the first and 
second line. $N=4000$ and $U=1.00$ and $t < 30000$. 
In figure IEP(c), the straight line is the average energy $<V_s> \approx 3.694$
of the separatrix and the dashed one the average energy $<V_M> \approx 3.958$ 
of the maximal value of the potential.} 
\label{figlevyener}
\end{figure}

\begin{figure} [h]
\psfig{figure=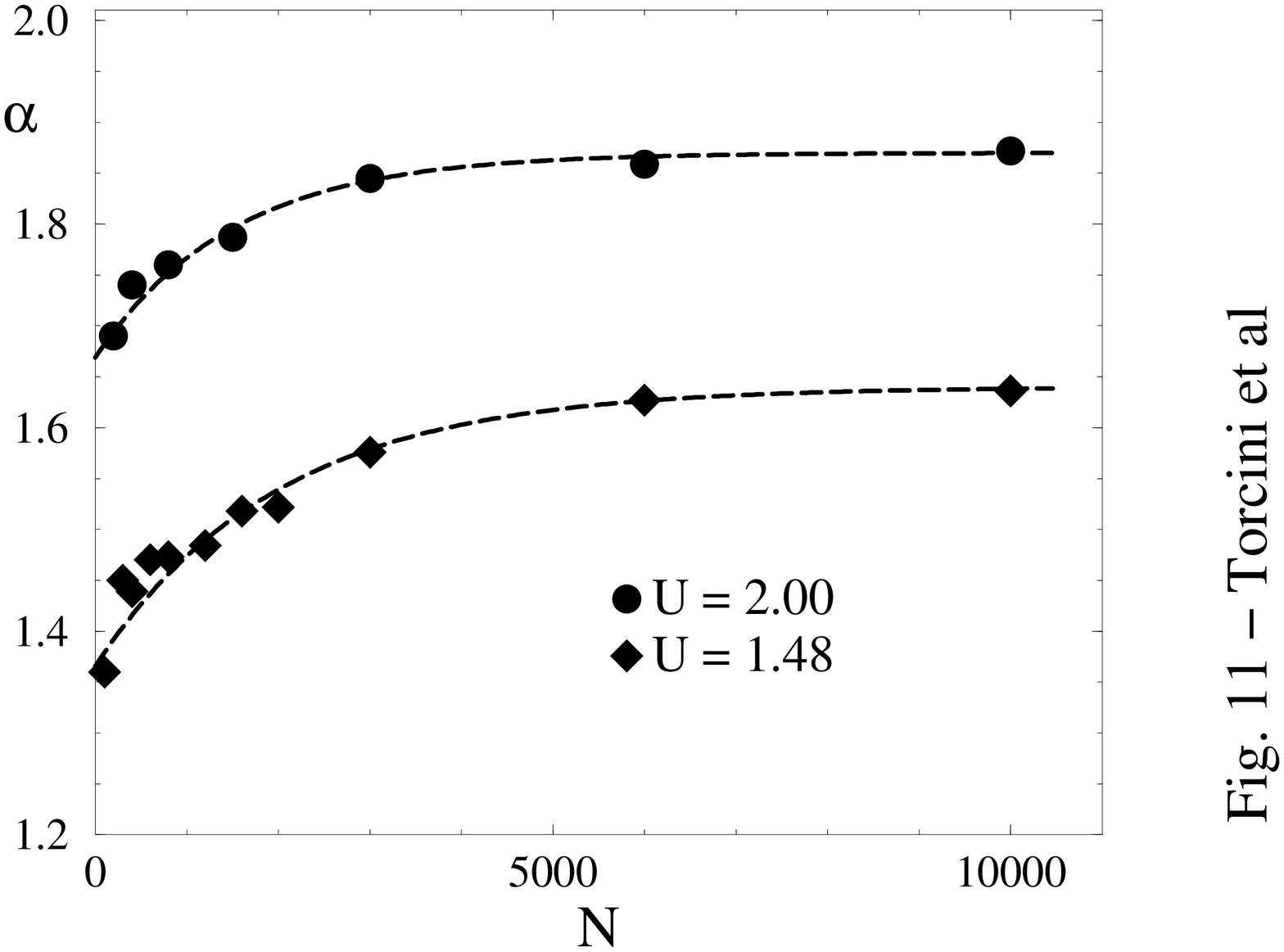,angle=-90,height=7truecm,width=7truecm}
\vskip 0.5truecm
\caption{Exponent $\alpha$ as a function of $N$ for $U=1.48$ 
(diamonds) and $U=2.00$ (circles). Dashed lines correspond to exponential
fitting. The data have been obtained for a total 
integration time $t=1.2 \times 10^6$. }
\label{fig10}
\end{figure}

\begin{figure} [h]
\psfig{figure=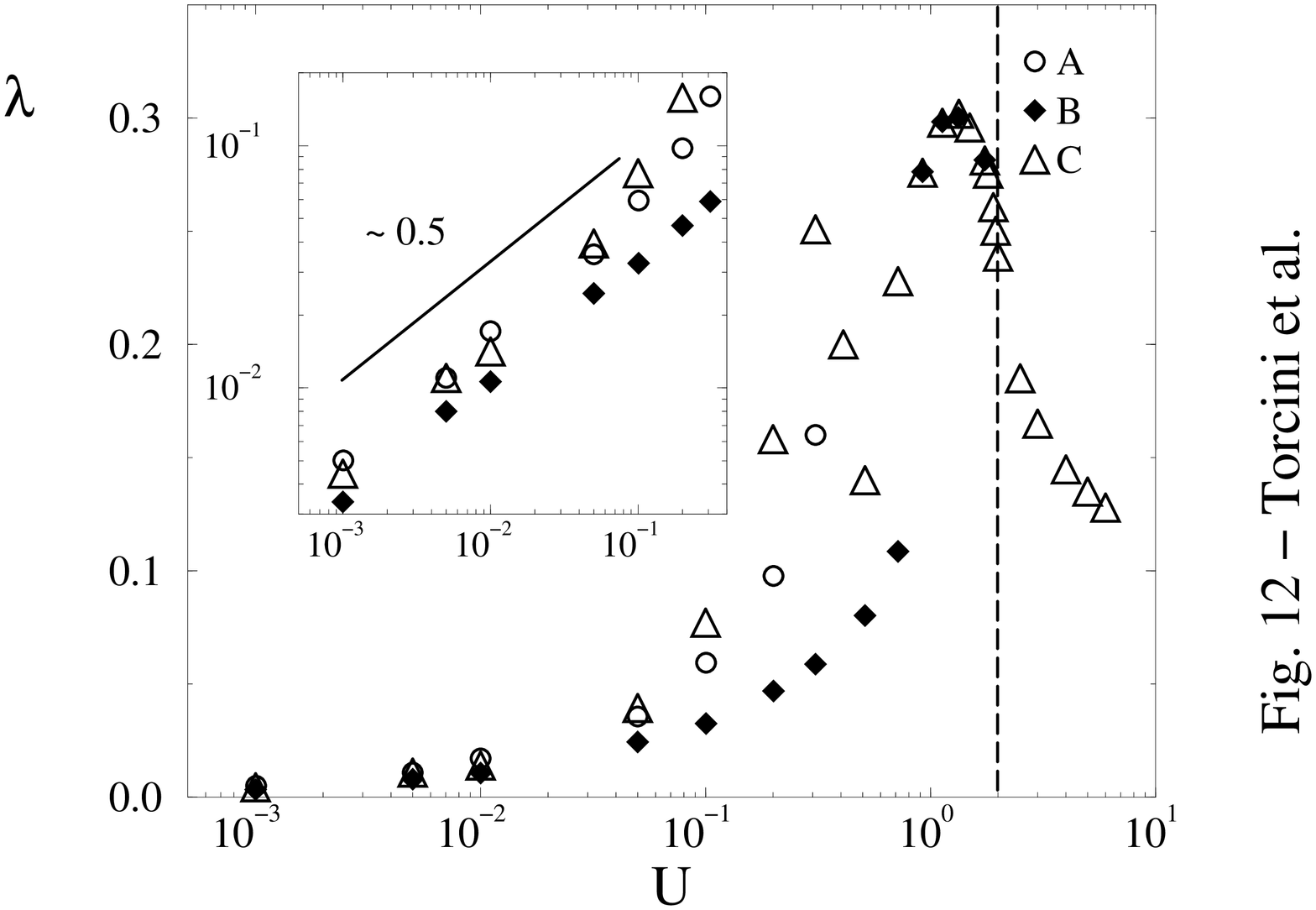,angle=-90,height=7truecm,width=7truecm}
\vskip 0.5truecm
\caption{Maximal Lyapunov exponent as a function of $U$ in a 
log-linear plot, for three different types of initial conditions 
(for details see the text). The dashed line indicates the critical
energy $U_c$.
In the inset, $\log(\lambda)$ versus $\log(U)$ is reported
at low energy. A scaling $\lambda \propto U^{1/2}$ is evident
in the low energy limit for all the three initial conditions.
All the reported data corresponds to $N=200$ and to integration times 
$1 \times 10^6 \le t \le 9 \times 10^6$.} 
\label{fig11}
\end{figure}

\begin{figure} [h]
\psfig{figure=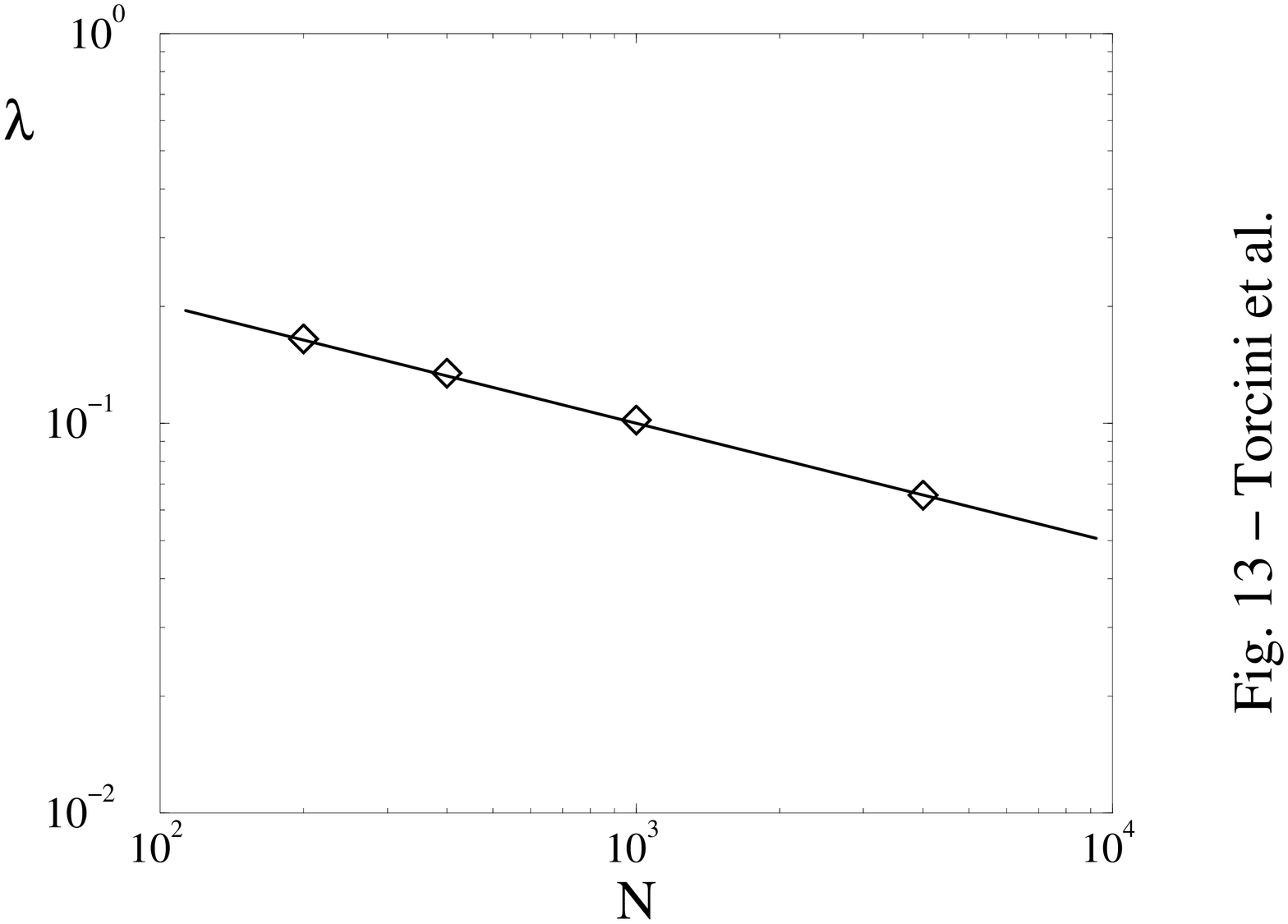,angle=-90,height=7truecm,width=7truecm}
\vskip 0.5truecm
\caption{$\log(\lambda)$ versus $\log(N)$ for $U=3.0 > U_c$.
The straight line is a linear fit to the data with slope $0.31$.
} 
\label{fig12}
\end{figure}

\begin{figure} [h]
\psfig{figure=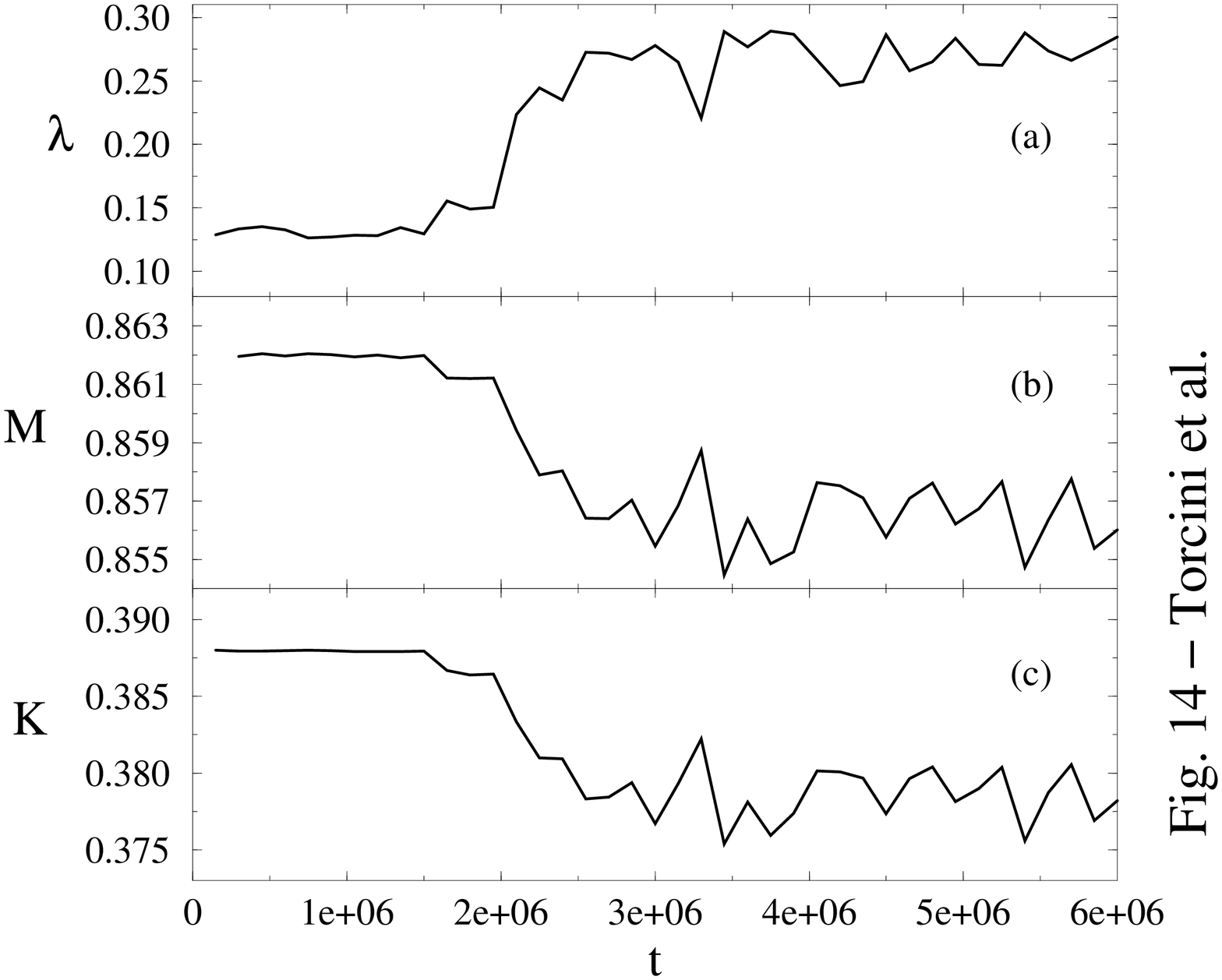,angle=-90,height=7truecm,width=7truecm}
\vskip 0.5truecm
\caption{Time evolution of the maximal Lyapunov exponent, of the 
magnetization $M$ and of the kinetic energy $K$ are reported for 
an initial condition of type (B) for $U=0.87$ and $N=200$.
} 
\label{fig13}
\end{figure}

\begin{figure} [h]
\psfig{figure=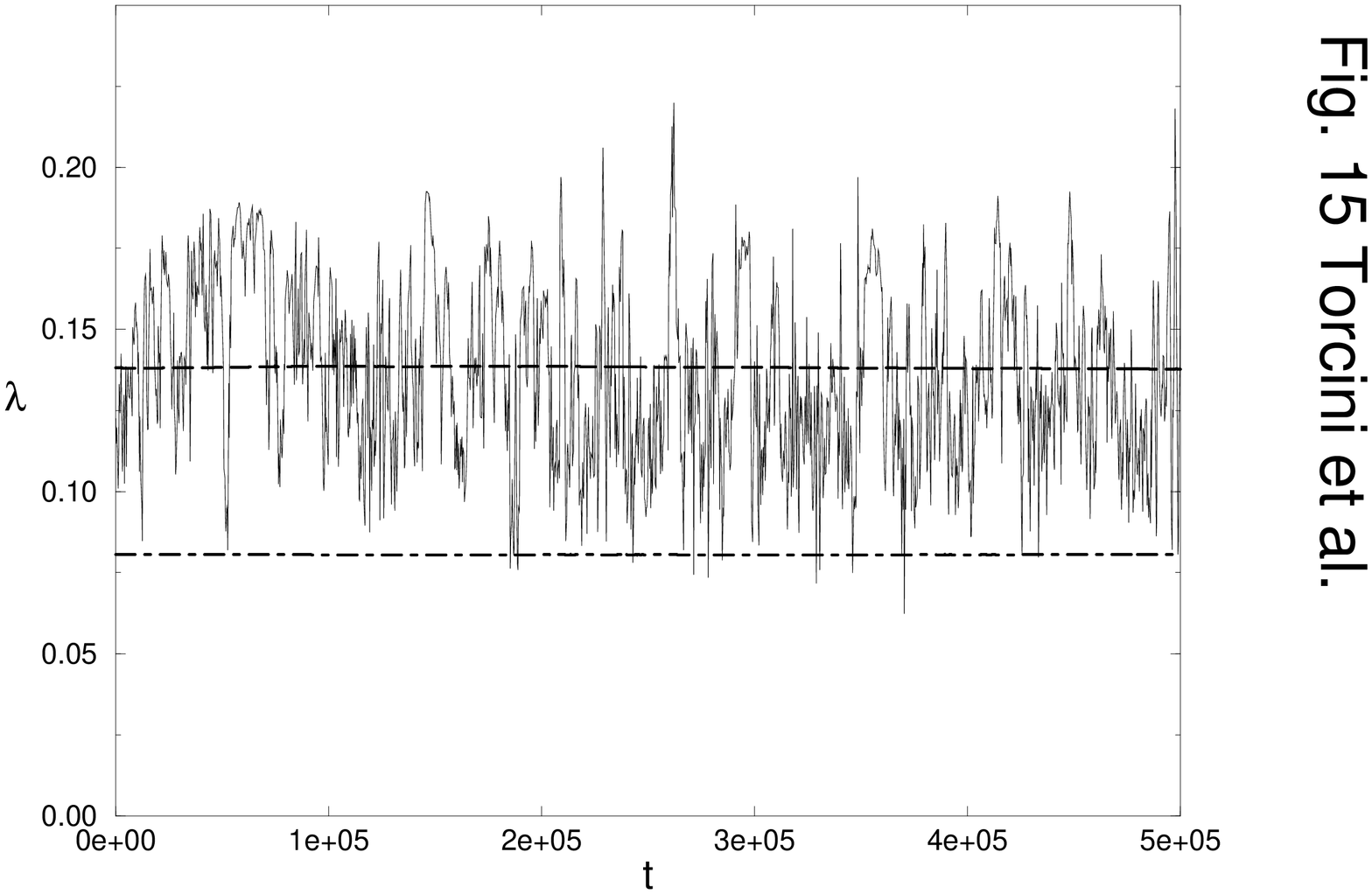,angle=-90,height=7truecm,width=7truecm}
\vskip 0.5truecm
\caption{The maximal Lyapunov exponent averaged over
short times ($t = 300$) is reported for an initial 
condition of type (C) (solid line) together with the
corresponding running average (long-dashed line).
The lower dot-dashed line corresponds to the 
running average of $\lambda$ for an initial
condition (C). The energy for both cases is $U=0.5$ 
and $N=200$.
} 
\label{fig14}
\end{figure}


\begin{thebibliography}{99}
\bibitem{note} It should be remarked that the diffusion mechanism
of the clustered particles induced by the separatrix crossing of
some of them is very fast when compared with the time scale
necessary to LEP's to escape from the potential well. As we 
have already said such escaping times are longer than our
simulation times. However, even if the particles inside the
cluster remain trapped, they can visit a phase space domain
wider than the one they could visit for a fixed potential
landscape.
\end{thebibliography}
\end{document}